\documentclass[useAMS,usenatbib]{mn2e}
\usepackage{ textcomp }
\include{journal_abbv}
\usepackage{graphicx,natbib}	
\usepackage{amsmath}	
\usepackage{amssymb}	
\usepackage{multicol}
\usepackage{multirow}
\usepackage{bm}		
\usepackage{pdflscape}	
\usepackage{float}
\usepackage[colorlinks=true,citecolor=blue]{hyperref}
\usepackage{longtable}
\vbadness=\maxdimen

\newcommand{\apj}{ApJ}
\newcommand{\apjs}{ApJS}
\newcommand{\apjl}{ApJL}
\newcommand{\aap}{A{\&}A}

\newcommand{\mnras}{MNRAS}
\newcommand{\aj}{AJ}
\newcommand{\araa}{ARA\&A}
\newcommand{\pasp}{PASP}

\newcommand{\nat}{Nature}
\newcommand{\ssr}{Space Science Reviews}
\newcommand{\pasa}{PASA}

\title[Dual AGN in DPAGN Galaxies]
  {Searching for Dual AGN in Galaxies with Double-Peaked Emission Line Spectra using Radio Observations}
\author[Rubinur et al.]{{K. Rubinur}$^{1,2}$\thanks{E-mail: rubinur@iiap.res.in}, M. Das$^1$,  P. Kharb$^3$ \\
$^1$Indian Institute of Astrophysics, Koramangala II Block, Bangalore 560034, India\\
$^2$Pondicherry University, R. Venkataraman Nagar, Kalapet, Pondicherry 605014, India\\
$^3$National Centre for Radio Astrophysics - Tata Institute of Fundamental Research, S. P. Pune University Campus,\\ Ganeshkhind, Pune 411007, India}

\begin{document}
\label{firstpage}
\maketitle
\begin{abstract}
Supermassive black hole (SMBH) binaries form due to galaxy mergers and minor accretion events. When the SMBHs are accreting, they form dual or binary AGN and can give rise to double-peaked emission lines in the optical spectra of the merger remnant. The double-peaked emission lines could also be due to jet-ISM interaction or rotating disks. One of the best ways to confirm dual/binary AGN in double-peaked AGN (DPAGN) is by using high resolution radio observations. We have observed a sample of 20 DPAGN at two or more frequencies using the Karl G. Jansky Very Large Array (VLA), of which one source is already published and the remaining 19 are presented in this paper. We have detected dual radio structures at separation of $\lesssim$ 10 kpc in three of our sample galaxies. Using the spectral index maps and optical spectra of the sources, we have confirmed that one of them is a dual AGN (DAGN), while the other two can be dual AGN or AGN+ star-forming nuclei pairs. Of the remaining sources, one has a clear core-jet structure and another source could be a core-jet structure or a DAGN. The remaining 13 sources are single cores while one source is not detected at any frequency. We find that for our dual AGN detection, the DPAGN emission lines do not originate from the dual/binary AGN. Instead, they could be due to outflows or jets. Hence, we conclude that DPAGN identified in low resolution SDSS spectra are not good indicators of dual/binary AGN. On the other hand, closely interacting galaxies or merger remnants are good candidates for detecting dual/binary AGN.
\end{abstract}


\section{Introduction} \label{section1}
In the merger driven picture of galaxy evolution, as galaxies merge, their supermassive black holes (SMBHs) lose angular momentum and spiral in towards the center of the merger remnant \citep{begelman.etal.1980,mayer.etal.2007,Komossa2016}. Cosmological theory \citep{Kulier2015} and simulations \citep{Menou2001,Roskar2015} also result in the formation of binary SMBHs. Mergers can trigger gas accretion onto the SMBHs, leading to the formation of active galactic nuclei (AGN) pairs or sometimes AGN-star forming nuclei pairs. At observed separations of $\sim$0.1 to 10~kpc, the AGN pairs are generally referred to as dual AGN (DAGN). At closer separations of a few times 10~pc or less, the SMBHs become gravitationally bound and form SMBH binaries or a binary AGN. In this stage, the stars are ejected from the surrounding region via gravitational scattering \citep{khan.etal.2011} until finally, the SMBH orbit shrinks through the emission of gravitational radiation and the SMBHs coalesce \citep{berczik.etal.2006,holley-bockelmann.khan.2015}. This gravitational radiation can be detected using pulsar timing array \citep[PTA,][]{Manchester2013}, e-Laser Interferometer Space Antenna \citep[eLISA,][]{eLISA2013}. Binary/dual AGN can produce significant feedback, both positive and negative which can have considerable effects on the merging process. It can also affect the galaxy environment through star-formation and AGN related outflows \citep{Koss2012,Blecha2013,Mezcua2014}.

According to models of galaxy mergers \citep{begelman.etal.1980,Milosavljevic2001,Yu2002,Merritt2013}, DAGN should be fairly common but studies show that confirmed DAGN are still relatively rare \citep{Das2017, Rubinur2018}. One of the reasons for the low detection rate is that the resolution required to resolve the individual AGN in dual or binary systems is high, usually sub-arcsec or milli-arcsec-scales for the nearby (z$\leq$1) Universe \citep{An2018}.
This resolution can be easily attained in high frequency radio interferometric observations such as with the Karl G. Jansky Very Large Array \citep[VLA,][]{Perley2011} where 8.4 GHz A-array can result in a resolution of 0.20$^{\prime\prime}$ and the Very Long Baseline Array (VLBA) where 8.4 GHz can provide a resolution of 1 mas. The very long baseline interferometry (VLBI) technique is essential to confirm the high brightness temperature, compact or extended emission from AGN on parsec- and sub-parsec-scales. The closest binary AGN have been detected with the VLBA \citep{Rodriguez2006,Kharb2017b}. However, the VLA with its higher sensitivity on sub-arcsec-scales is best to search for kpc-scale dual AGN \citep{Burke2014,Fu2015,mullersanchez2015}. Radio observations have the additional advantage of penetrating dust.

There could be other reasons too for the rare detection - such as rapid binary coalescences timescales \citep{dotti.etal.2015}, where the SMBH mergers are accelerated by external factors such as gas accretion or interactions with passing massive objects. To understand the evolution of SMBH pairs from separations of few kpc to final coalescence, we need a larger sample of DAGN.

\citet{Zhou.etal.2004} suggested that double-peaked emission lines in AGN spectra can be an effective way to detect DAGN when two AGN have individual narrow line regions (NLRs). Blue-shifted and red-shifted components of the NLR lines will produce double-peaked emission lines in the spectra. Double-peaked emission lines have been detected in AGN spectra for the last $\sim$50 years \citep{Sargent1972,Veilleux1991,Heckman.etal.2004} and were thought to be due to outflows or rotating disks. However, they can also arise from NLR gas kinematics \citep{Gelderman1994,Stockton2007,Fu2009,Kharb15}. High resolution Hubble Space Telescope (HST\footnote{www.spacetelescope.org/}) observations and imaging with Adaptive Optics (AO) have revealed that there are many galaxies (e.g NGC 1068) where the double-peaks in emission lines arise from the interaction of jets/outflows with the NLR \citep{Crenshaw.etal.2000}. The double-peaked emission lines can also be due to nuclear rotating ionized gas disks \citep{muller.etal.2011,Kharb15}.

About $\sim$1\% of all low redshift AGN identified by the Sloan Digital Sky Survey (SDSS) display double-peaked [O III]$\lambda$5007 emission \citep{Smith.etal.2010}. As double-peaked lines can arise from NLR kinematics, jet-ISM interaction or outflows, one needs to carry out high resolution X-ray or radio imaging to confirm the presence of DAGN. Many DAGN samples have been made from SDSS DPAGN galaxies and high resolution radio, optical, IR or in X-ray observations have been done to search for DAGN \citep[e.g.,][]{wang.etal.2009,ge.etal.2012,Shen2011,Comerford2012,Liu2010}. Of all these different approaches, high resolution radio interferometry has the distinct advantage that it is not affected by dust obscuration and can give sub-parsec resolution images of the radio emission from two AGN. The disadvantage is that not all AGN have enough radio emission to be mapped at high resolutions since only 10\% of galaxies are radio loud. At X-ray wavelengths, {\it Chandra} has the highest resolution (0.5$^{\prime\prime}$) which is comparable to the resolution obtained from VLA observations. But again, not all AGN show X-ray emission. 

Double-peaked broad emission lines have also been thought to be the tracer of binary/dual AGN \citep{Gaskell1983,Boroson2009}. However, this signature can also be explained by a Keplerian accretion disk \citep{Eracleous2003}. Confirming the presence of binary/dual AGN in broad double-peaked emission line galaxies is difficult because of the large outflow signature in the optical spectrum. Instead, very high spectroscopic resolution and long term monitoring is required to measure the systematic changes due to orbital motion \citep{fu.etal.2012}. 

Apart from double-peaked emission lines, there are other indirect signatures of binary/dual AGN, such as S- or X- shaped radio galaxies \citep[e.g.,][]{Ekers.etal.1978,begelman.etal.1980,Rottmann.etal.2001,Hodges.2012,Nandi2017,Rubinur2017} or periodicity in optical variability \citep[e.g.,][]{Lehto.etal.1996,graham.etal.2015,Liu.etal.2015}. We have discussed this extensively in our earlier paper \citep{Rubinur2017}. 

In this paper, we present high resolution radio observations of 19 DPAGN
using the Karl G. Jansky Very Large Array (VLA). Part of the sample were  
visually selected as disky galaxies and the remaining galaxies had non-disky morphologies. 
In section \ref{section2}, we describe our method of sample selection. In section \ref{section3}, we discuss our VLA observations and the archival data used in this study.
Data reduction process is discussed in section \ref{section4}. The results are discussed in section \ref{section5};
the radio and the spectral index images are described in section \ref{section5.1} while the SMBH masses, accretion rates and 
star formation rates (SFRs) are calculated in section \ref{section5.2}. 
We discuss the implications of our results in section \ref{section6} and the conclusions are summarized in section \ref{section7}. 
Throughout this paper, we assume a flat $\Lambda$CDM cosmology model with parameters $H_{0}=~73.0$~km~s$^{-1}$~Mpc$^{-1}$  , $\Omega_{m}=~0.27$ and $\Omega_{\Lambda}=~0.73$. The spectral index, $\alpha$, is defined such that the flux density at frequency $\nu$ is S$_\nu\propto~\nu^\alpha$.

\begin{figure}
\includegraphics[width=\columnwidth]{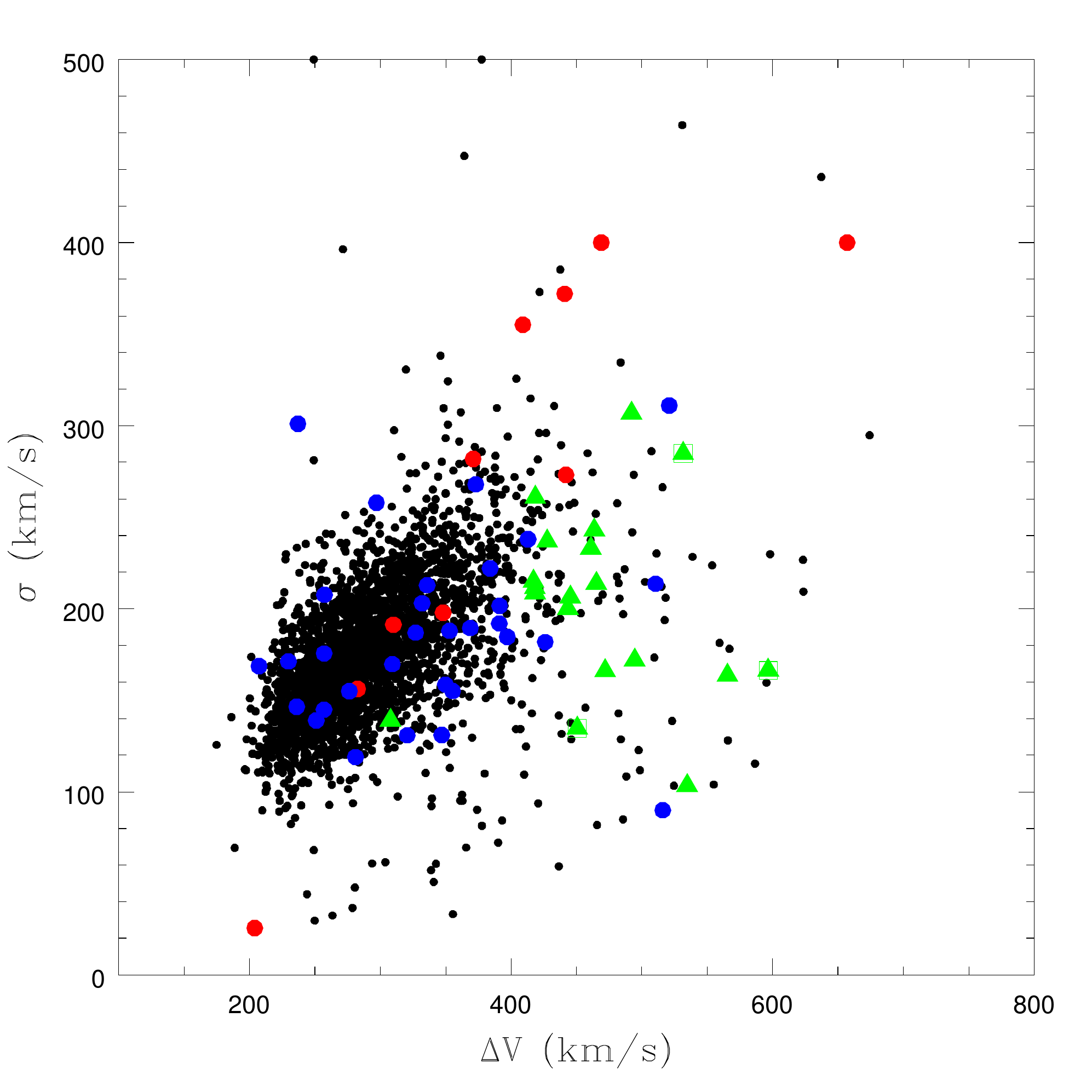}
\label{delv_sig}
\caption{\small The $\sigma-\Delta V$ plot of 3030 double-peaked
emission line galaxies from \citet{ge.etal.2012}. The velocity ($\Delta$V) is
the Doppler separation between the double-peaks in the [OIII] emission lines.
$\sigma$ is the nuclear velocity dispersion derived from the bulge dispersion.
The \citet{ge.etal.2012} points are in black
and our nineteen target sample galaxies are in green. The
red points are all confirmed dual AGN which have SDSS spectra
and the blue points are the DPAGN which did not
yield the dual AGN in follow-up high resolution observations
(From literatures; Table \ref{conf_dagn}).}
\end{figure}

\section{Sample selection}\label{section2}

We started with an initial sample of 3030 DPAGN from \citet{ge.etal.2012}, which was mainly
composed of nuclei that had double peaked [O III] narrow emission lines (NEL) of separation $\Delta V$. Out of 3030 DPAGN, only 81 are type 1 AGN and hence have broad line components. 
The NEL nature of the remaining sample meant that they did not have prominent AGN outflows. 
The double peaks in [O III] can be due to dual/binary AGN, jet- ISM interaction or rotating nuclear disks. In order to distinguish between DAGN and rotating disks, we plotted in Figure \ref{delv_sig}, the velocity difference 
between the [O III] emission lines ($\Delta V$), and the stellar bulge velocity dispersion 
($\sigma$) for the 3030 NEL sample of \citet{ge.etal.2012}. There is a significant correlation between $\Delta V$ and $\sigma$ (Spearman rank correlation coefficient = 0.57, probability $<10^{-5}$). This implies that these data lie on
a line or plane, depending on whether the coefficient of proportionality is
constant or not for all the sample DPAGN.
The $\sigma \propto\Delta V$ correlation is likely due to a
rotating disk as explained below.

For gas rotating in a disk of mean radius R around an SMBH of mass $M_{BH}$ the velocity $v$ is given by, 
\begin{equation}
 GM_{BH}/R^{2} = (v^2/R) 
\end{equation}
We assume that the virial theorem holds for the nuclear region
so that 2$T_E+V_E$=0, where $T_E$ is the mean total kinetic energy and $V_E$ is the
mean total potential energy of stars moving with a velocity dispersion
$\sigma$ around a central SMBH of mass $M_{BH}$ in a region of radius $R$.
We can then use the relation
\begin{equation}
 GM_{BH}/R = 2({\sigma}^2/2) 
\end{equation}

Using these two equations, we obtain $\sigma\propto v$. If the double peaked [OIII] line 
is due to a rotating disk, then we can assume $v= \Delta V$/2. Note that the rotating disks may be inclined at various angles with respect to our line of sight, so the general trend follows the $\sigma\propto v$ relation. Also, $\sigma$ is the nuclear bulge velocity dispersion derived from the underlying absorption lines in the optical SDSS\footnote{www.sdss.org/dr12/} spectrum. So, assuming that points along the $\sigma\propto v$ correlation in the $\sigma~vs~\Delta V$ plot represent mainly rotating disks, we have tried to select DPAGN that are offset 
from this relation. However, this simple analysis cannot separate the jet-ISM interaction cases from the DAGN/rotating disk ones. 
 
In Figure \ref{delv_sig}, we have over-plotted confirmed DAGN from
the literature for which $\sigma~vs~\Delta V$ data are available (red
points, Figure \ref{delv_sig}; Table \ref{conf_dagn}). We have also over-plotted the DPAGN that did not yield
confirmed DAGN in follow-up high resolution observations (blue points, Figure \ref{delv_sig}; Table \ref{conf_dagn}). We find that most of the
confirmed DAGN have $\Delta V>400$~km s$^{-1}$ while the non-confirmed DPAGN have $\Delta V<400$~km s$^{-1}$. This can be roughly explained
as being due to the increase in relative velocities between the SMBHs as
they come closer; hence DPAGN with larger $\Delta V$
are more likely to be dual AGN, provided the DPAGN emission is not associated
with a rotating disk or jet-cloud interaction.
Using the above two conditions we chose DPAGN from Figure \ref{delv_sig} that (i)~do not lie on the 
$\sigma$~$\propto~\Delta V$ correlation and (ii)~have $\Delta V$~$\geq$~400~km~s$^{-1}$. It must be noted
that these approximations are not based on any rigorous calculations and the $\sigma\propto v$
plot was constructed to increase our chance of DAGN detections. In fact, since we had first made 
the plot (year 2014), there have been three detections in the $\sigma\propto v$ region of 
Figure \ref{delv_sig} (Table \ref{conf_dagn}). However, most of the detected DAGN are still offset from this region.  

In the following subsections, we discuss how the samples were selected using the $\sigma~vs~\Delta V$
plot. In sample~1, we restricted the sources to be low redshift disky galaxies that had $\Delta$V~$\geq$~400~km~s$^{-1}$. In sample~2 however, we included all the galaxies, irrespective of their morphologies, that are offset from the $\sigma\propto v$ correlation and had $\Delta V$~$\geq$~400~km~s$^{-1}$.

\subsection{Sample~1}\label{section2.1}
Most of the detected dual AGN are present in elliptical galaxies.
Dual AGN in spiral or disk galaxies are relatively unknown, probably because they represent minor mergers and may be difficult to distinguish in distant galaxies. In fact there are only 3 examples of DAGN in a disk galaxy. Earlier NGC 3393 \citep{Fabbiano2011} was the only spiral galaxy which was known to be a DAGN. SDSS J113126.08$-$020459.2 \citep{Shangguan2016} and NGC 7674 \citep{Kharb2017b} are two recent detection/candidate of dual/binary AGN in spiral galaxies.
Hence in this study,although our broad goal was to detect DAGN in galaxies, we first focused on searching for DAGN in spirals as there are very few detections in the literature. We chose the galaxies from the $\sigma-\Delta V$ plot that were (i)~offset from the $\sigma\propto v$ line; (ii)~$\Delta V$ $>$ 400 km~s$^{-1}$, (iii)~z $<$ 0.1, (iv)~clear double peaked [O III] lines (since [O III] comes from the NLR and is a better dual AGN indicator), (v)~showed clear disk morphologies (by visual inspection) in their SDSS images and (vi)~had radio emission in the NRAO VLA Sky Survey (NVSS\footnote{https://www.cv.nrao.edu/nvss/}; \citep{Condon1998}) or Faint Images of the Radio Sky at Twenty cm (FIRST\footnote{sundog.stsci.edu/}; \citep{Becker1994}). 

We rejected galaxies that had been studied in earlier radio observations \citep[e.g., UGC 4229:][]{Nagar1999}. We finally obtained six galaxies based on the aforementioned criteria. We also chose two more disk galaxies from \citet{fu.etal.2012} which have clear double-peaked H$\alpha$ and [O III] lines and show radio emission in NVSS and FIRST images. Both the galaxies have z $<$ 0.1 but one of them has $\Delta V$ = 307~km~s$^{-1}$. We observed these galaxies in VLA cycles 15A (6 GHz) and 16A (15~GHz). We have published one of the sources 2MASXJ12032061+1319316 in \citet{Rubinur2017} as it shows an interesting kpc scale S- shaped core-jet structure. Hence, in this paper we discuss only the seven galaxies from sample 1 (see Table \ref{sam_gal}). 

\subsection{Sample 2}\label{section2.2}
Sample 2 galaxies were also selected from \citet{ge.etal.2012}. The sample 2 sources have similar selection criteria as sample 1 sources except that there is no restriction on morphology or redshift.
Our sample 2 consists of 14 DPAGN of which 2 have redshifts z $<$0.1, 11 sources have 0.1$\leq$ z $\leq$0.5 and only one source has z $>$0.5. Some of the galaxies in this sample are elliptical. Two of the galaxies are merger systems, where the morphologies are not clear. Two of the objects had right ascension coordinates that could not fit within the observational schedule and hence could not be observed. We finally observed a total of 12 sources (Table \ref{sam_gal}) in the VLA cycle 16B (at 8.5 and 11.5~GHz).

\begin{figure*}
\hspace{-0.5cm} \includegraphics[width=1.1\columnwidth,trim= 0 0 0 40]{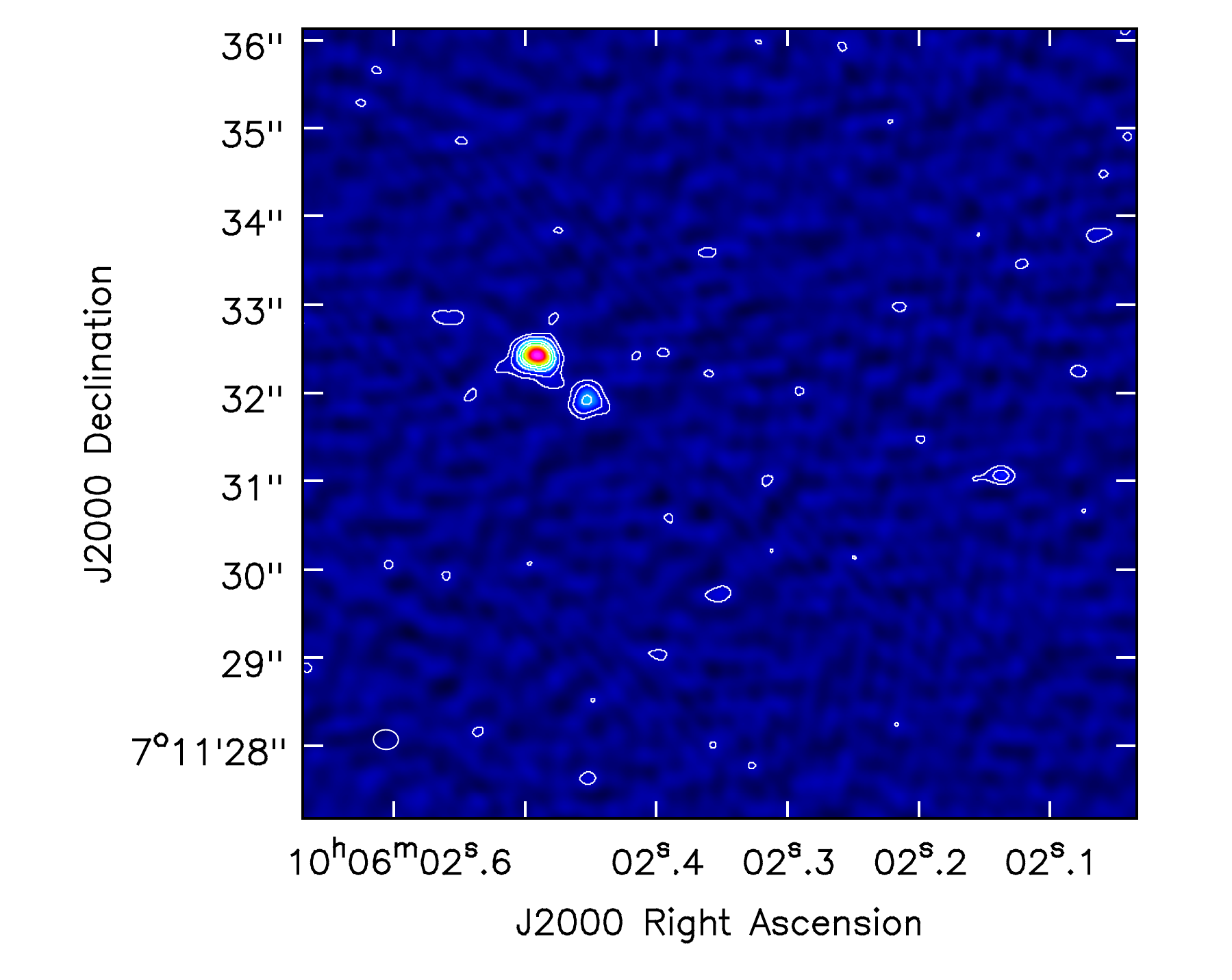}
\includegraphics[width=0.99\columnwidth]{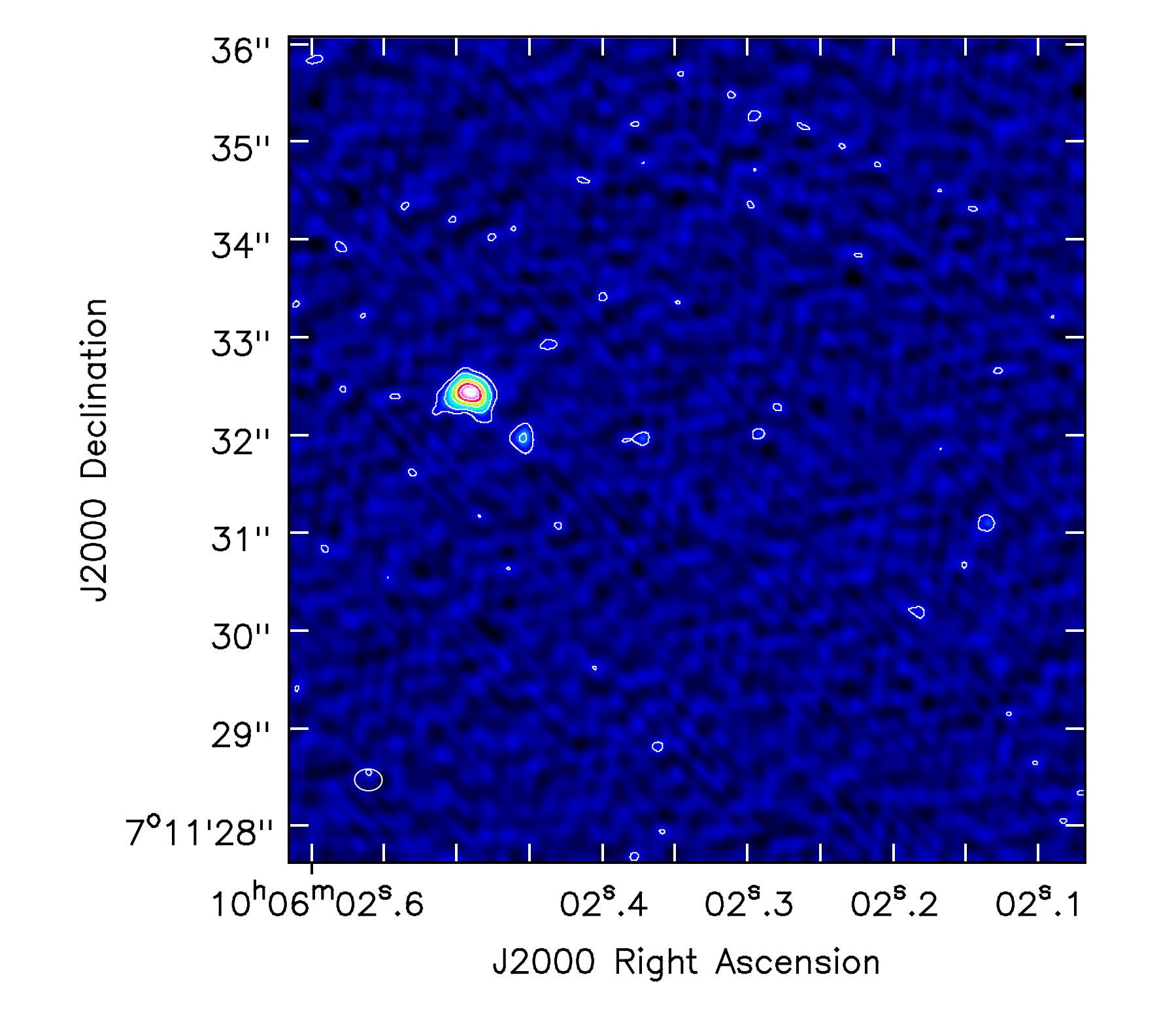}
\includegraphics[width=0.70\columnwidth,trim= 200 0 0 0]{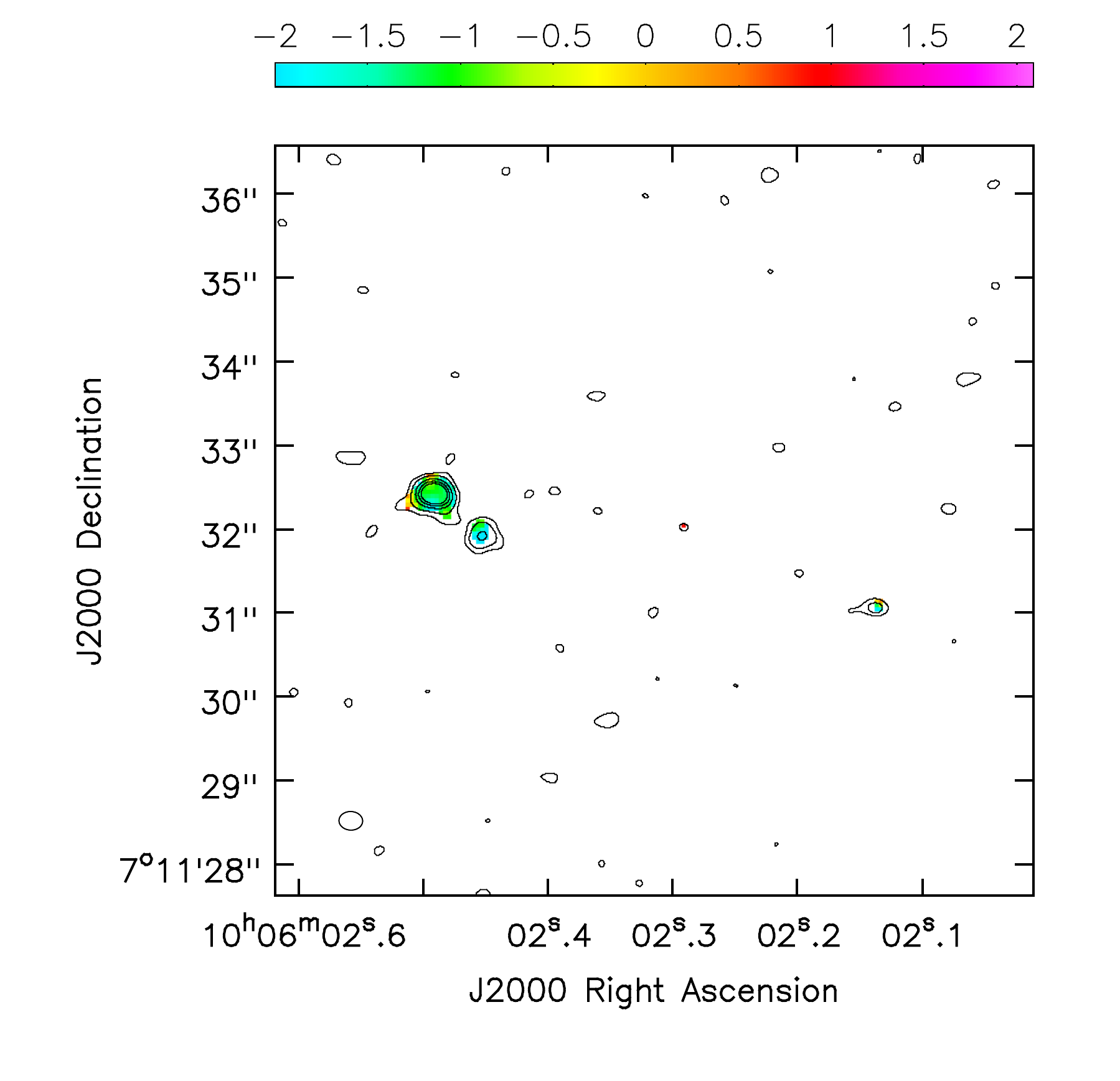}
\includegraphics[width=0.45\columnwidth, trim=00 -30 300 100]{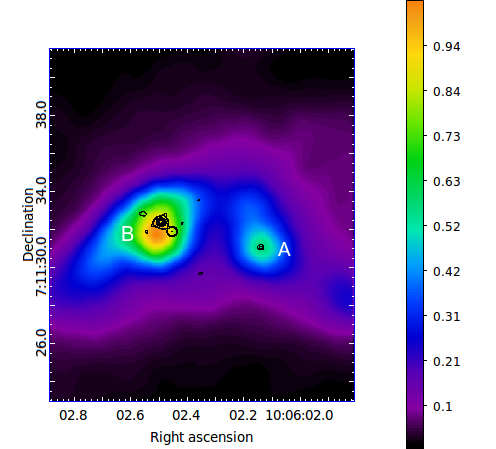}
\caption{The dual radio core galaxy J1006.
Moving from top left to right and bottom left to right : (i)~the radio image
at 8.5 GHz, (ii)~the radio image at 11.5 GHz, (iii)~the spectral index map using 8.5 GHz and
11.5 GHz images and (iv) the 8.5~GHz contours superimposed on the SDSS g band image of the
galaxy. This is a confirmed DAGN at the separation
of 12~kpc. The contour levels are 10\%, 20\%, 40\%, 60\% and 80\% of the peak intensities (table~\ref{dual_core}).}
\label{J1006}
\end{figure*}

\begin{figure*}
 \includegraphics[width=1.15\columnwidth]{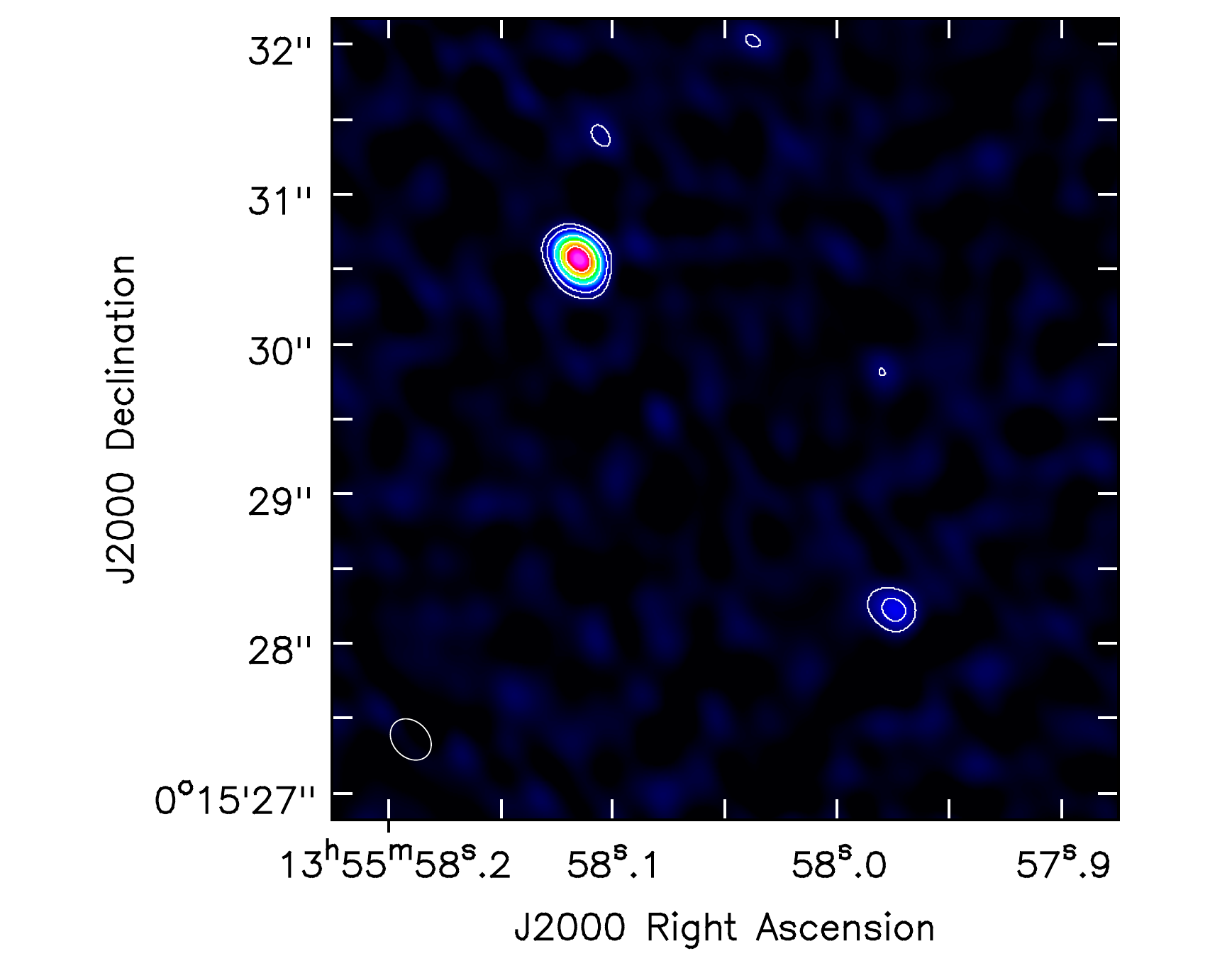}\includegraphics[width=1.1\columnwidth]{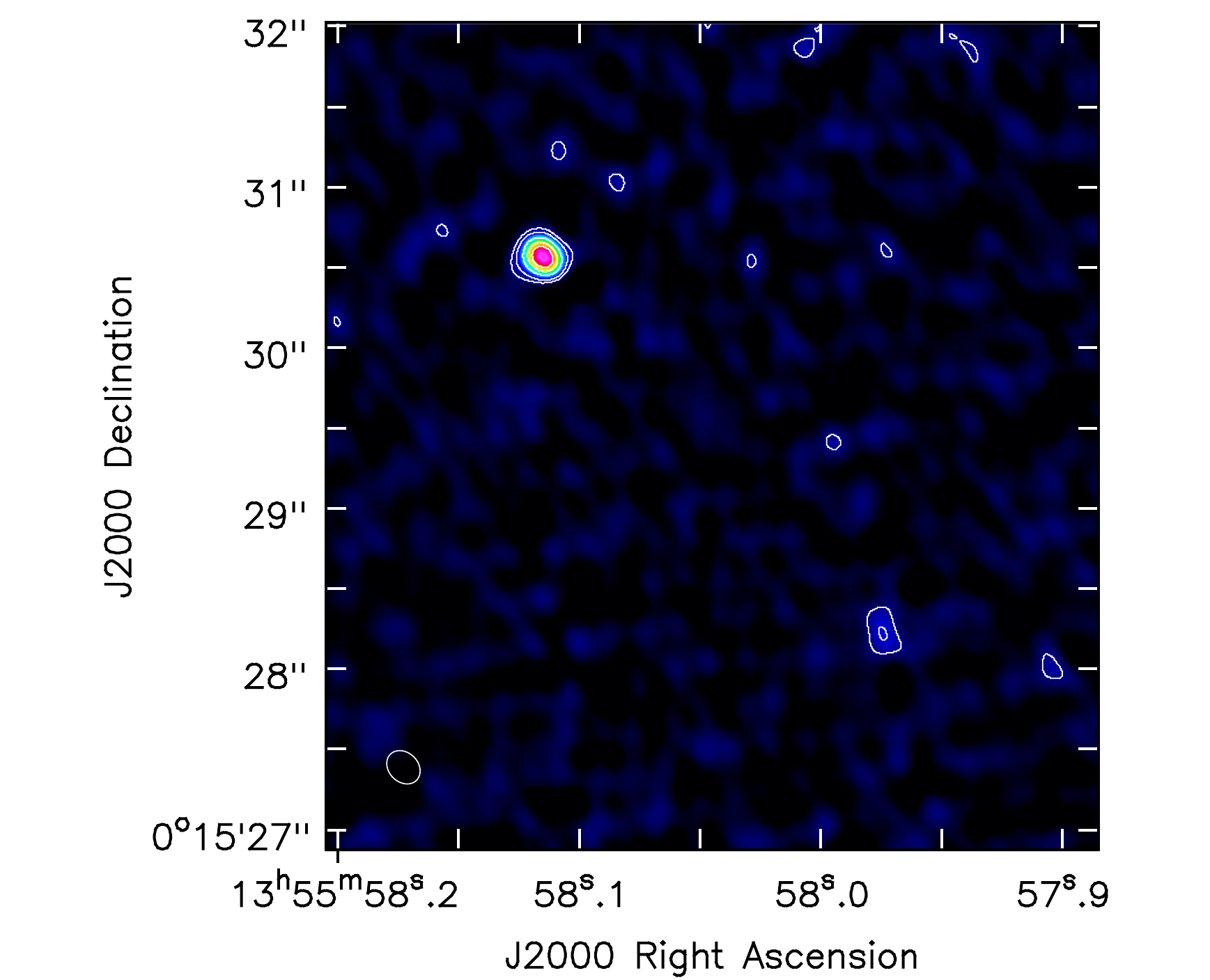}
 \includegraphics[width=.670\columnwidth,trim= 170 0 0 0]{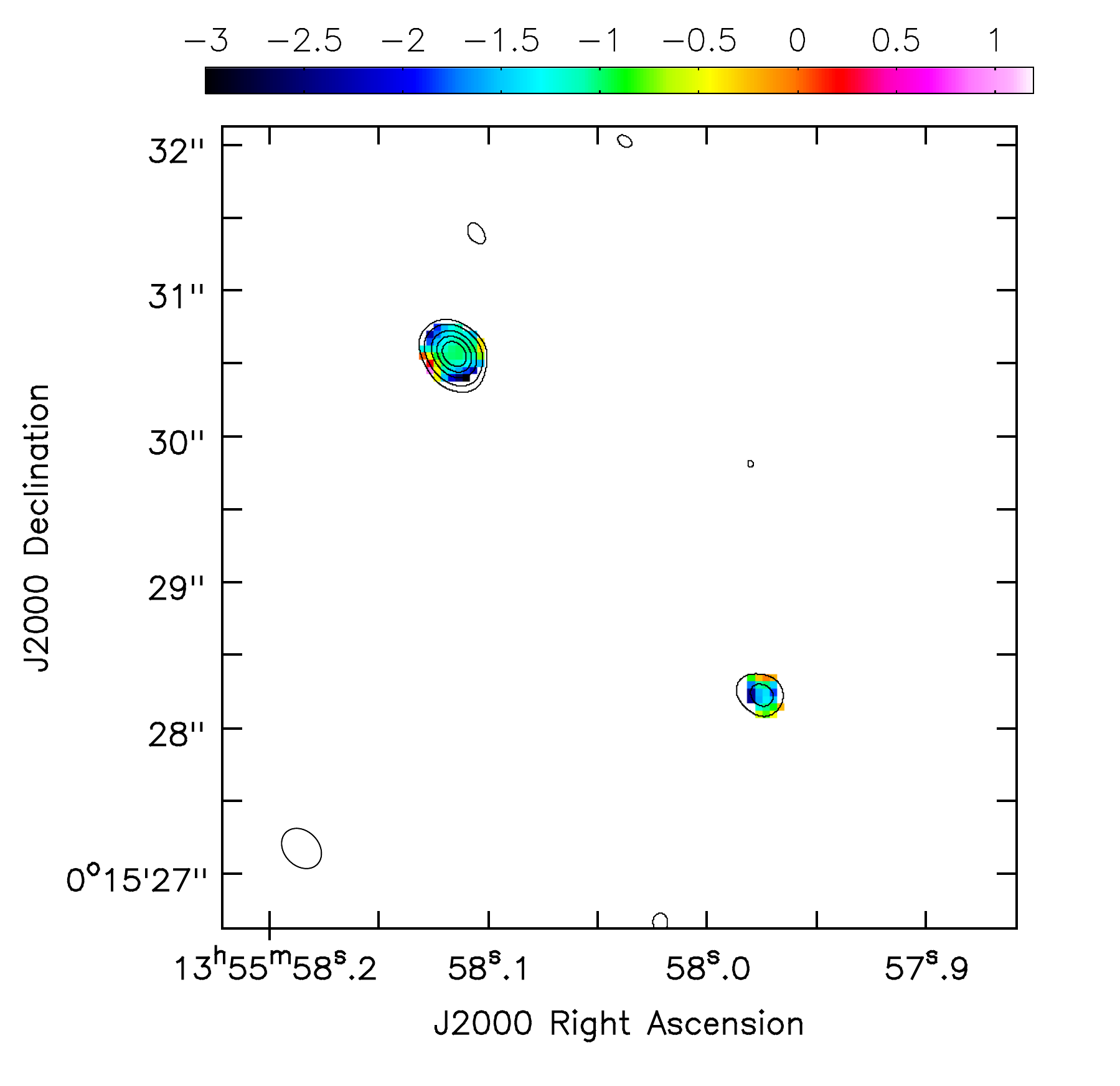}\includegraphics[width=.5\columnwidth,trim=0 0 300 600]{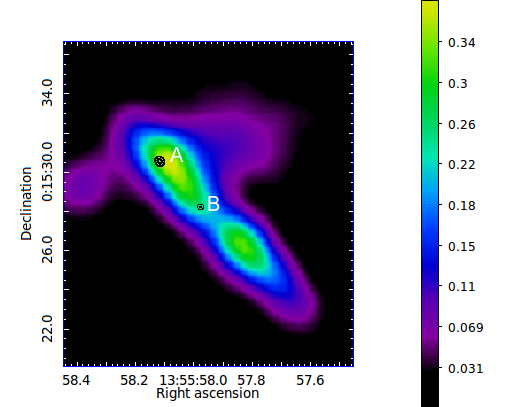}
\caption{The dual radio source galaxy J1355. Moving from top left to right and bottom left to right: (i)~the radio image at 8.5 GHz, (ii)~the radio image at 11.5 GHz, (iii)~the spectral index map using 8.5 GHz and 11.5 GHz images and (iv) the 8.5~GHz contours superimposed on the SDSS g band image of the galaxy. The contour levels are 10\%, 20\%, 40\%, 60\% and 80\% of the peak intensities (table~ \ref{dual_core}).}
\label{J1355}
\end{figure*}

\begin{figure*}
 \includegraphics[width=1.05\columnwidth]{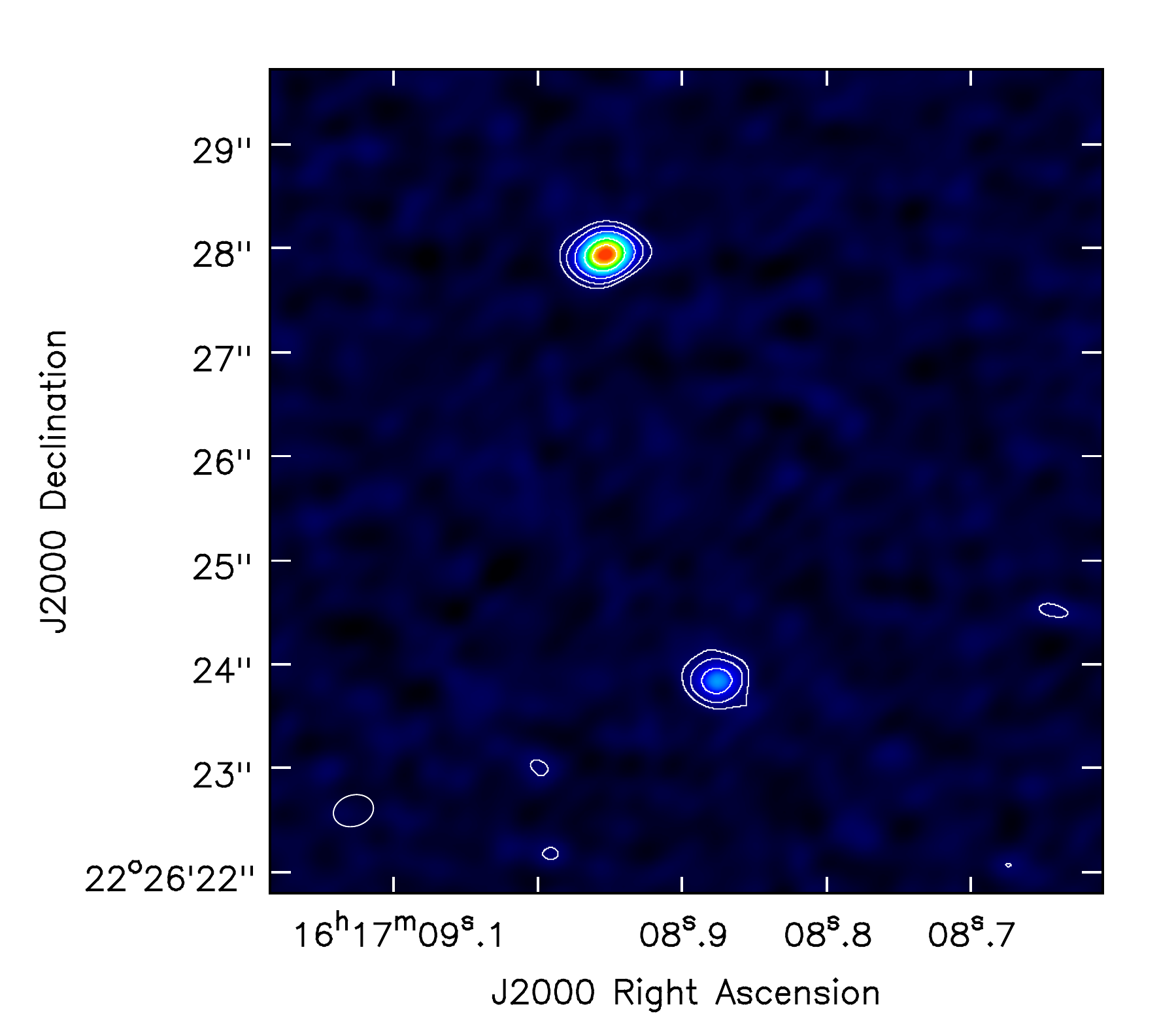}\includegraphics[width=1.07\columnwidth]{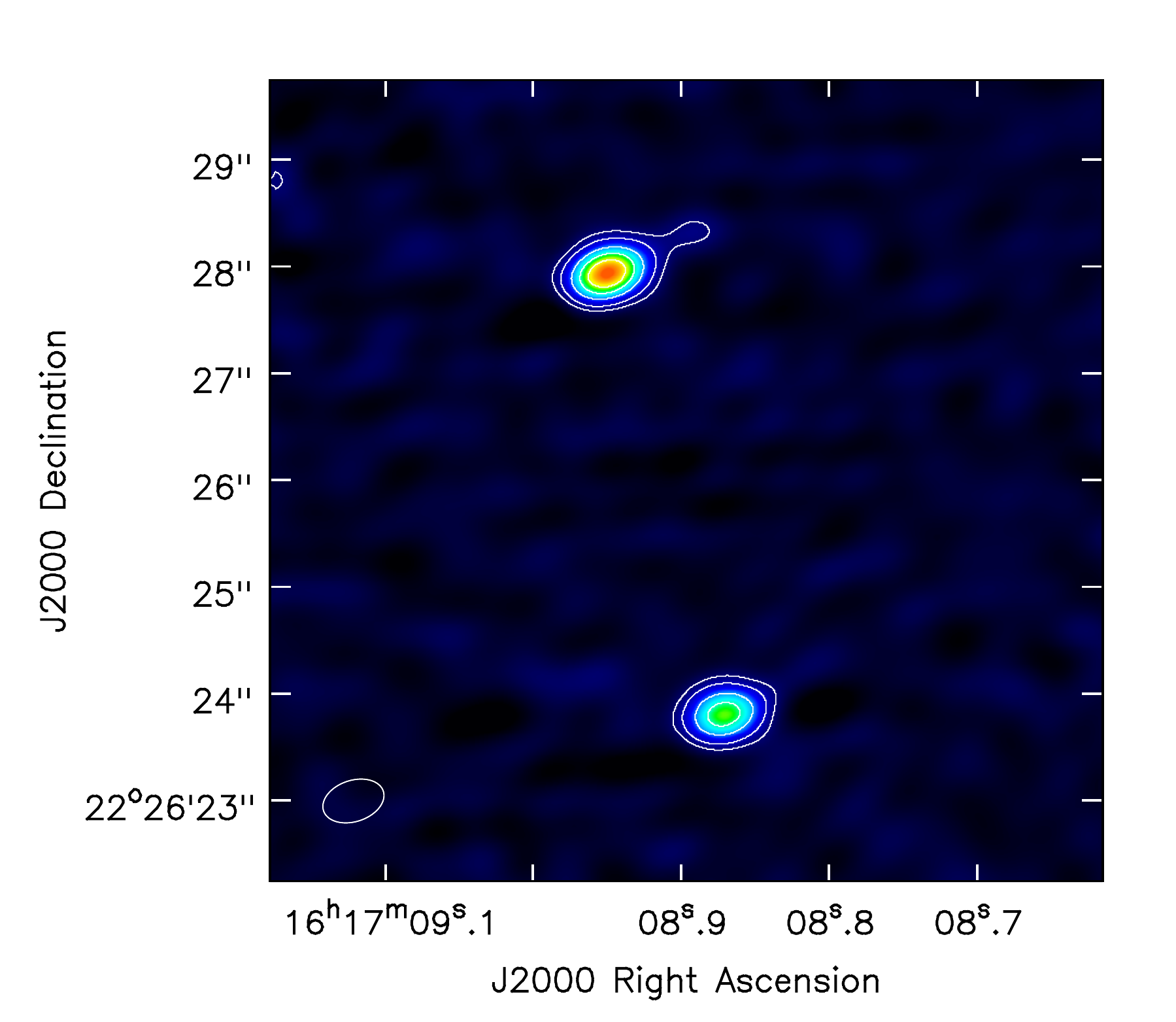}
 \includegraphics[width=0.85\columnwidth,trim= 150 0 0 0]{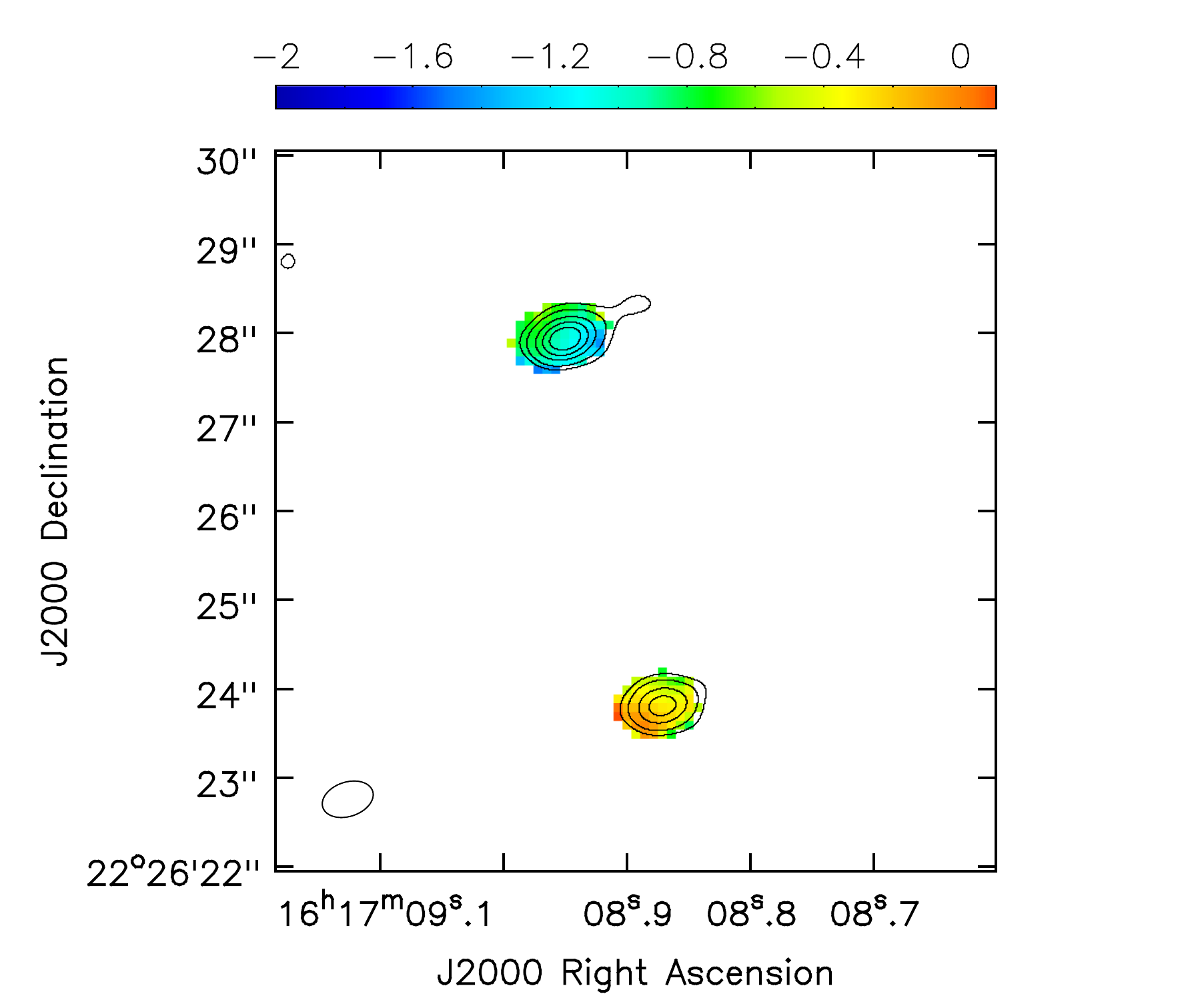}\includegraphics[width=0.9\columnwidth, trim=0 00 200 400]{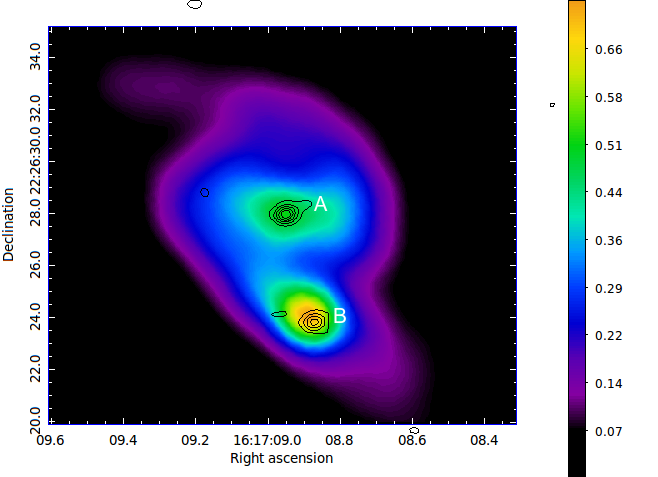}
\caption{The dual radio core galaxy J1617. Moving from top left to right and bottom left to right: (i)~the radio image at 6.0 GHz, (ii)~the radio image at 15.0 GHz, (iii)~the spectral index map using 6.0 GHz and 15.0 GHz images and (iv) the 15.0~GHz contours superimposed on the SDSS g band image of the galaxy. This can be DAGN or AGN+SF nuclei pair at the separation
of 5.6~kpc. The contour levels are 10\%, 20\%, 40\%, 60\% and 80\% of the peak intensities (table \ref{dual_core}).}
\label{J1617}
\end{figure*}

\begin{figure*}
 \includegraphics[width=0.68\columnwidth]{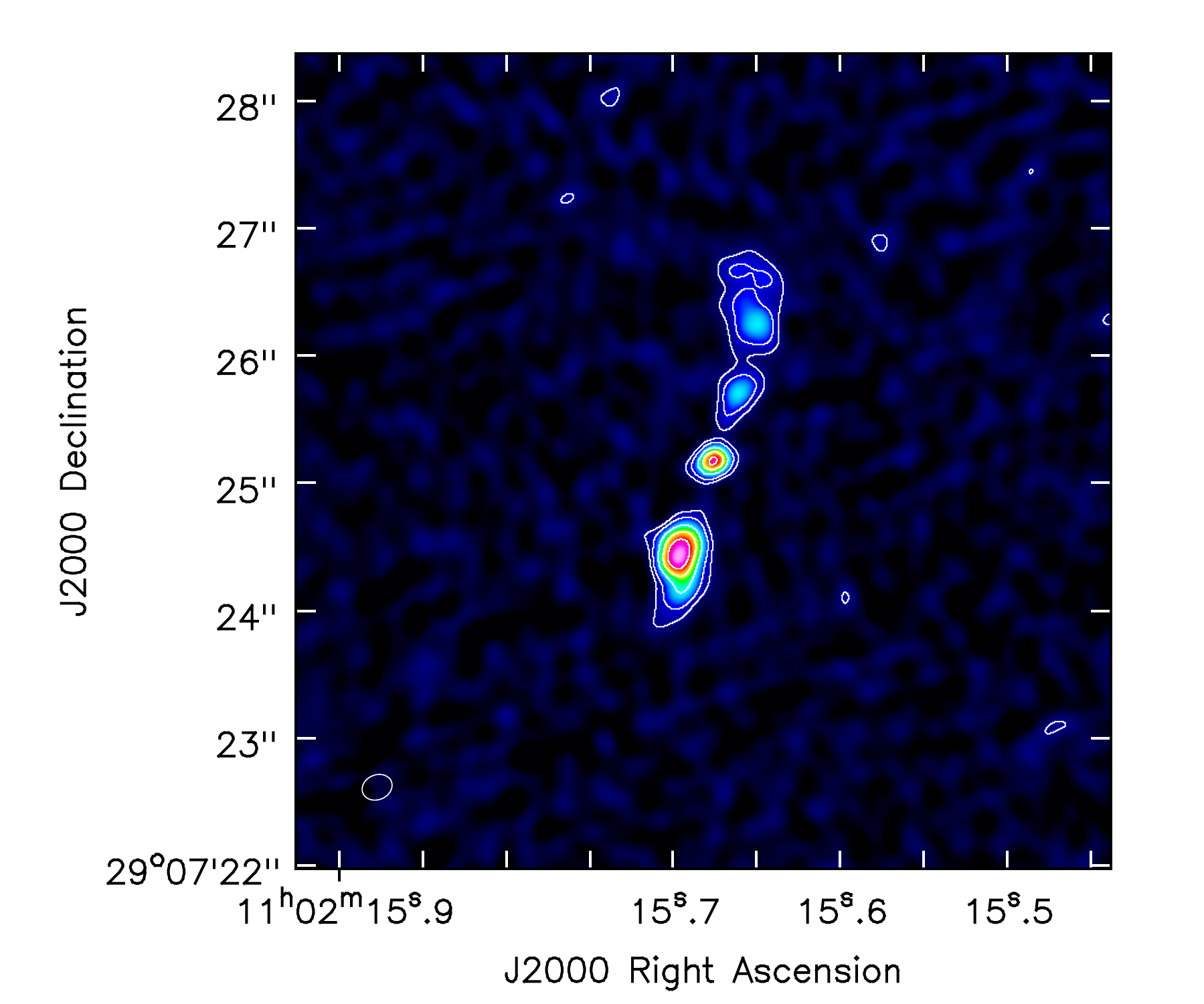}\includegraphics[width=0.70\columnwidth]{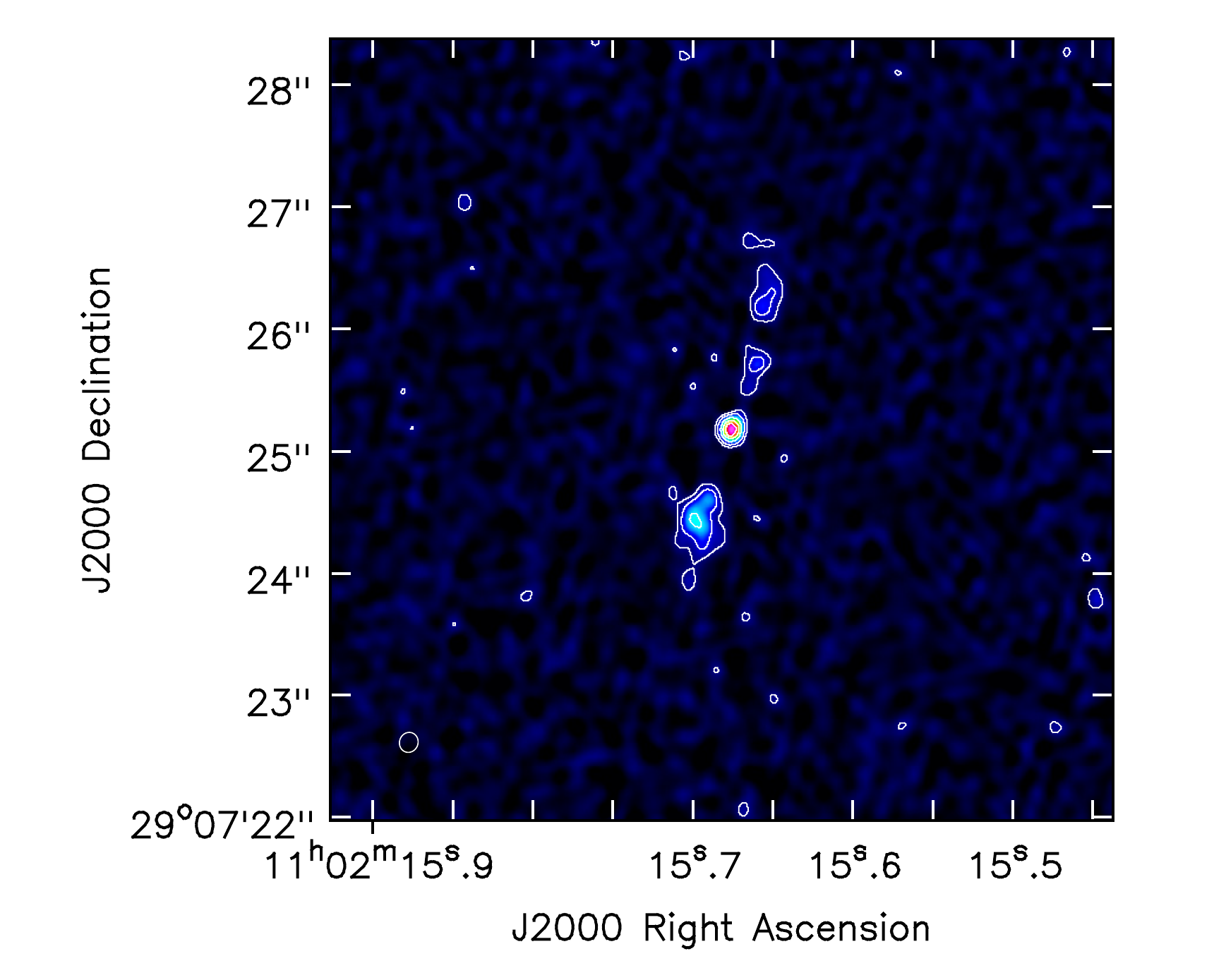}
 \includegraphics[width=0.65\columnwidth]{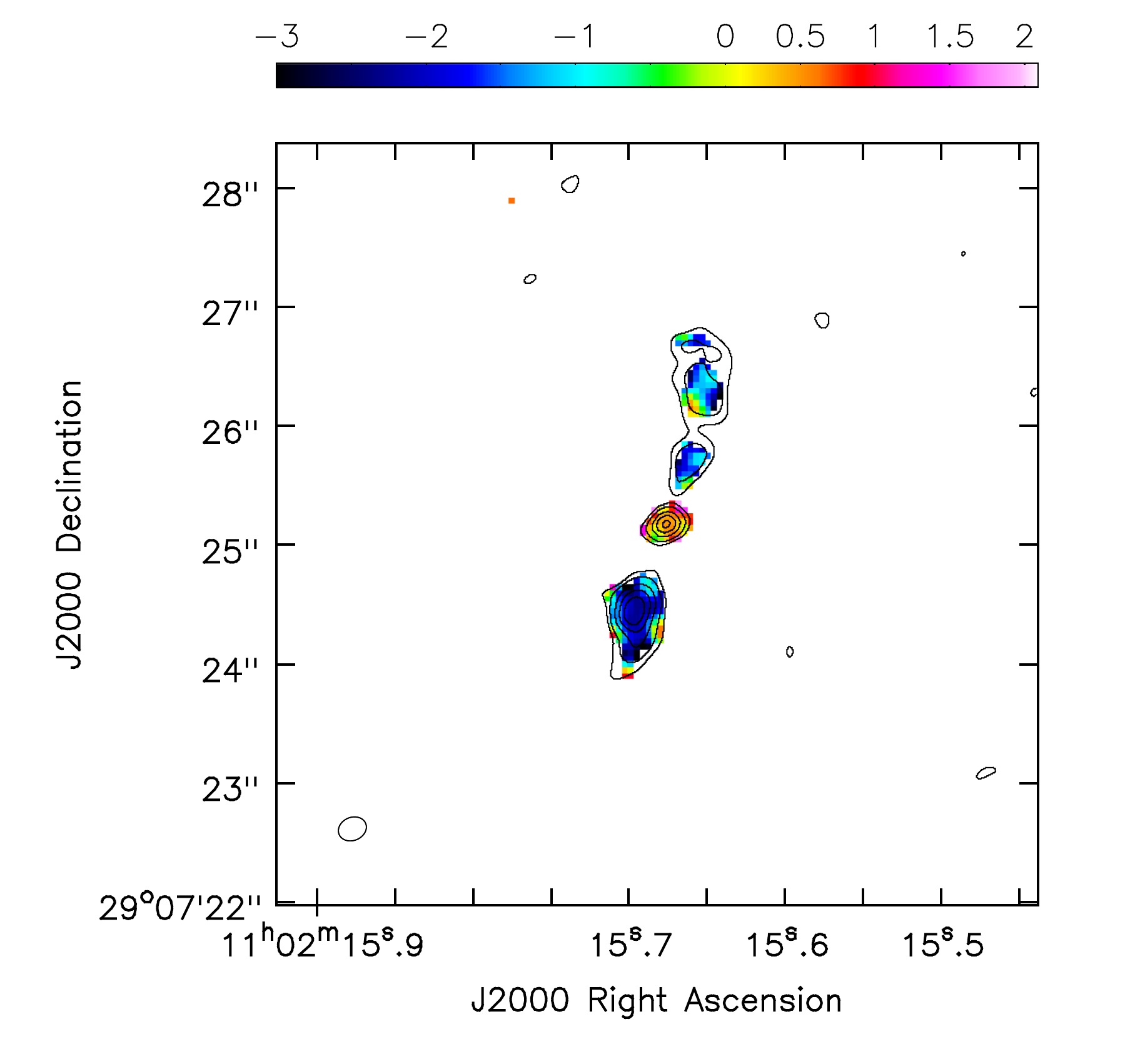}
\caption{The radio images and the spectral index map of J1102. Moving left to right is the 8.5 GHz image, 
11.5 GHz image and spectral index map using 8.5 GHz and 11.5 GHz images. The images are overlaid with its contours and the spectral index map 
is overlaid with the 11.5 GHz radio image contours. The contour levels are 10\%, 20\%, 40\%, 60\% and 80\% 
of its peak intensities respectively (table \ref{single_core}).}
\label{J1102}
\end{figure*}

\begin{figure*}
 \includegraphics[width=1.30\columnwidth,trim=60 0 80 0]{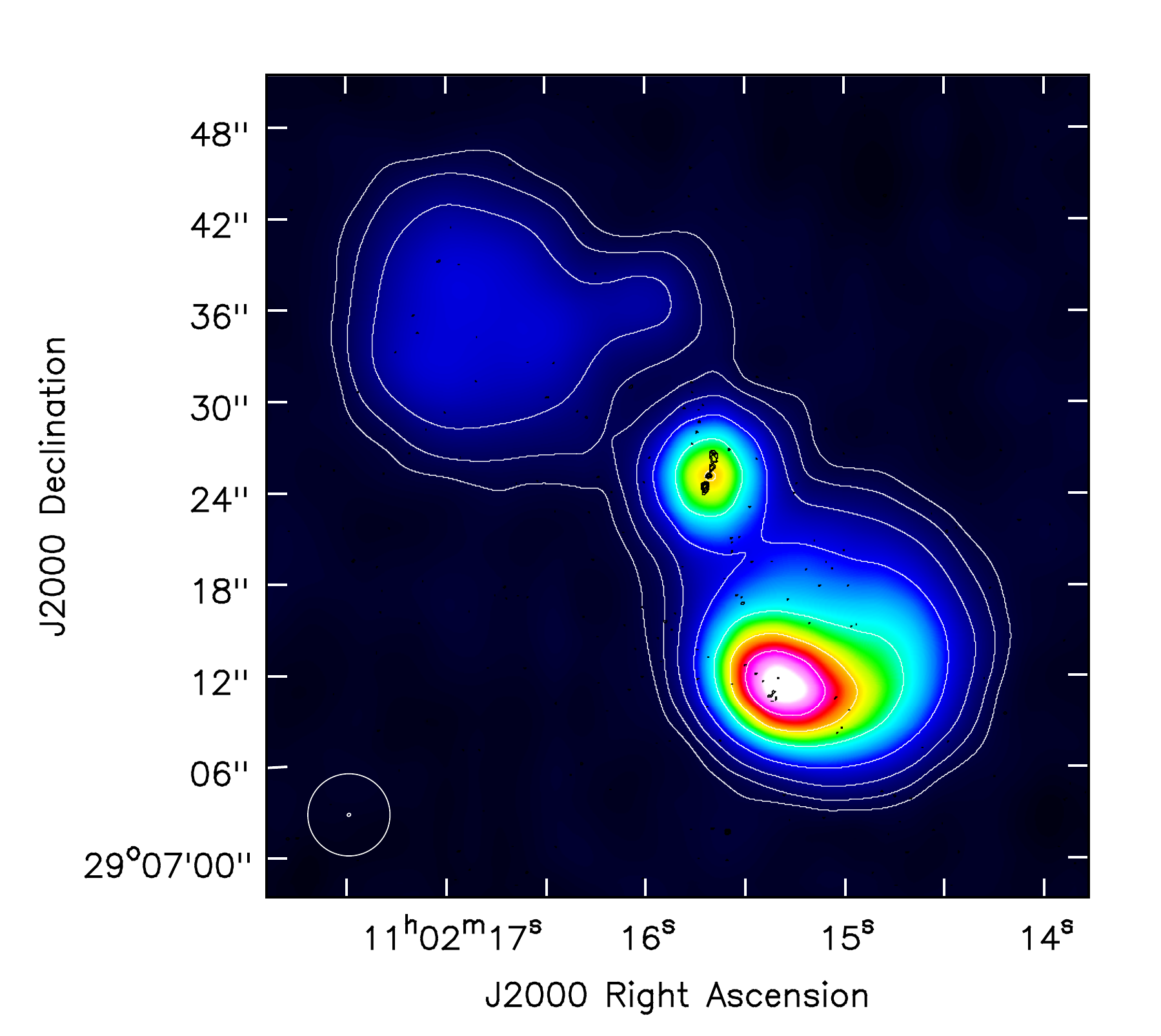}
\caption{The FIRST image with contours overlay with the 8.4 GHz image contours in black of J1102. 
The FIRST image shows $\sim$ 120 kpc jet and the 8.5 GHz shows ~6 kpc jet. There is change in the jet directions
from small scale to large scale. This can be a Z-shaped source.
The contour levels in FIRST image are 2.5\%, 5\%, 10\%, 20\%, 40\%, 60\% and 80\%  of its peak intensity respectively.
The contour levels in 8.5 GHz image are 10\%, 20\%, 40\%, 60\% and 80\%  of its peak intensity respectively (table \ref{single_core}).}
\label{J1102_FIRST}
\end{figure*}

\begin{figure*}
 \includegraphics[width=0.7\columnwidth]{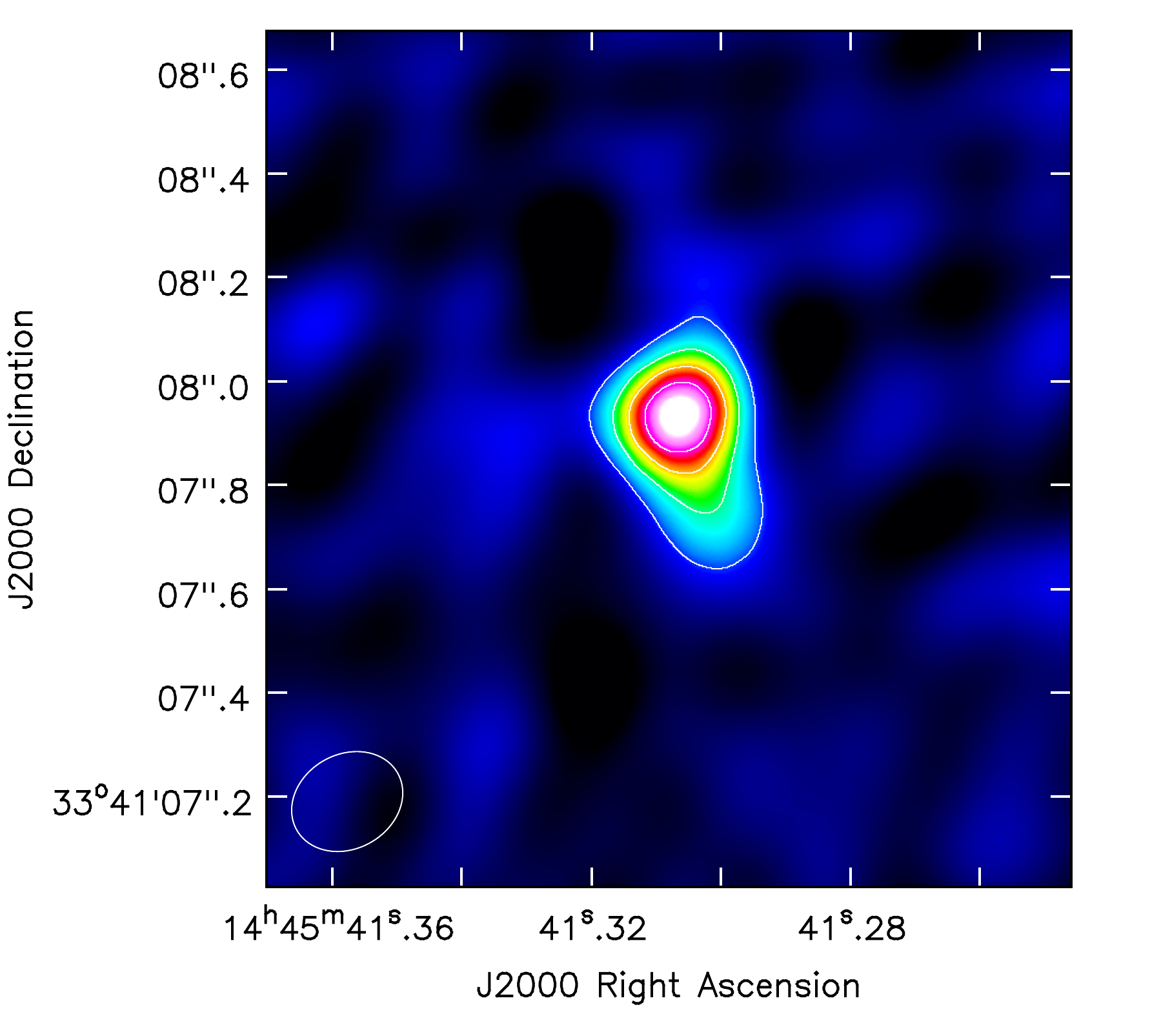}\includegraphics[width=0.77\columnwidth]{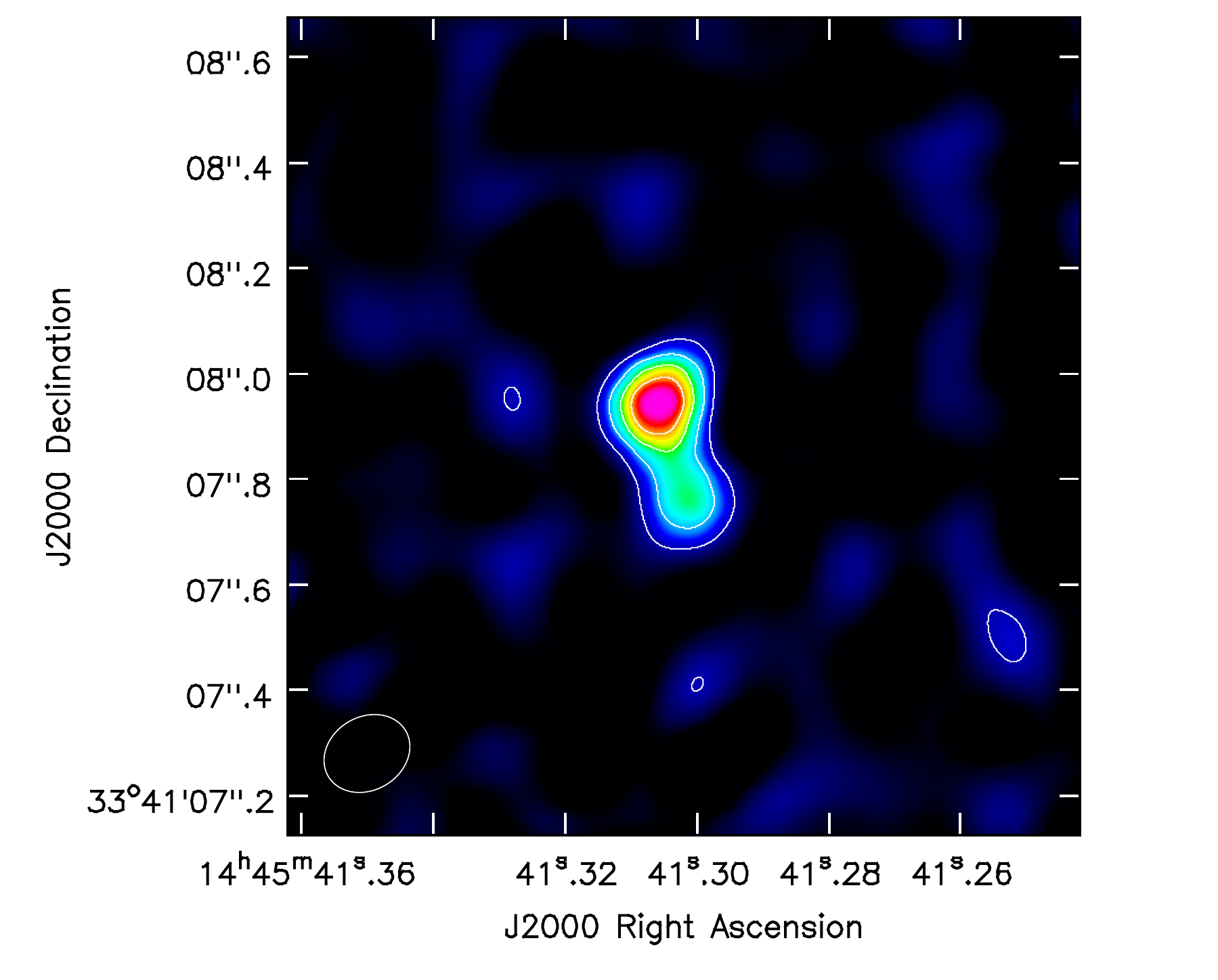}\includegraphics[width=0.73\columnwidth,trim=0 0 0 0]{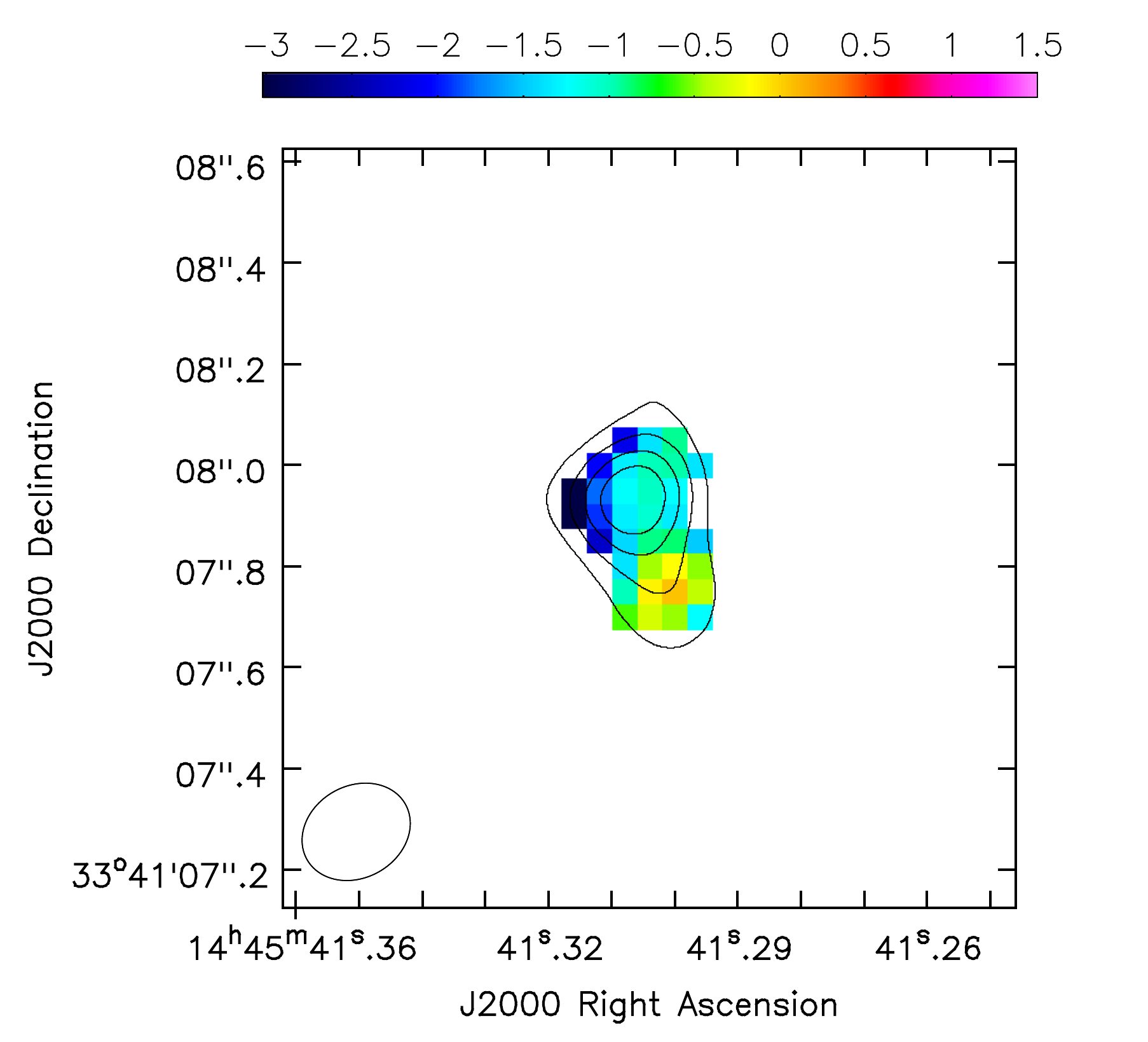}
\caption{The radio images with contours and the spectral index map of J1445. Moving from left to right is the 8.5 GHz image, 
11.5 GHz image and spectral index map using 8.5 GHz and 11.5 GHz images overlaid with the 11.5 GHz radio contours.
The contour levels are 20\%, 40\%, 60\% and 80\%  of its peak intensity respectively (table \ref{single_core}). 
J1445 shows an extended structure which can be core-jet or dual AGN.}
\label{J1445}
\end{figure*}

\section{New and archival data } \label{section3}
{$\bullet$ Sample 1:} Seven DPAGN were observed with the VLA 
at 6~GHz in the A-array configuration for 1.5 hours (Project ID: VLA/15A-068; Table \ref{obs_cat}). 
The observations were carried out with a 1792~MHz wide baseband centered at 5.935 GHz 
with 14 spectral windows (SPW), each window having 64 channels with a frequency resolution of 2 MHz. 
Each target source was observed for 4 minutes. 
Based on the preliminary results from the 6 GHz data, we planned to observe 5 out of 7 galaxies at 15~GHz, to obtain the spectral index images. We obtained 15 GHz VLA B-array data of 4 galaxies on 29 May 2016 (Project ID: VLA/16A-144; Table \ref{obs_cat}) because one galaxy already had high resolution VLA data available in the NRAO archive.
The 15 GHz observations were carried out with 1890 MHz baseband centered at 14.9 GHz. 
Each of the targets was observed for 15 mins. 
The fifth galaxy, J2304 had VLA archival data at 8.5-11.5 GHz (Project ID:VLA/13B-020).
J2304 was observed for 24 min at two central frequencies 8.5 GHz and 11.5 GHz.


{$\bullet$ Sample 2:} We observed 12 DPAGN in sample 2 with the VLA (Project ID: VLA/16B-002). 
The observations were done on 15 November 2016, at the two central frequencies of 8.5 and 11.5~GHz in A-array configuration with 1.6 GHz baseband at each frequency center with total of 16 spectral windows.  
The targets were observed for $\sim$6 minutes. The observational details are provided in Table \ref{obs_cat}.
List of the flux density and phase calibrators are given in Table \ref{calibrator}
The expected VLA resolution in the X band in the A-array configuration is 0.2$^{\prime\prime}$. The highest redshift of our sample galaxy is $z=0.355$ which gives a spatial scale of 7~kpc/$^{\prime\prime}$. So we can resolve up to 1.4~kpc in this galaxy with VLA observations, while SDSS can resolve only up to 21~kpc with its 3$^{\prime\prime}$ fiber.

\section{ Data Reduction} \label{section4}
The Common Astronomy Software Applications (CASA) \citep{McMullin2007} 
and  Astronomical Image Processing System (AIPS) \citep{Moorsel1996} packages were used for the VLA data reduction. First, we flagged the bad data of the calibrators. Bad data include RFI (radio frequency interference), zero amplitude data, shadowed antennas.  Once the RFI was identified, that particular frequency range or time was flagged interactively. We searched for any zero amplitude data by zooming in the {\tt plotms} display and then identified the corresponding antenna and time range. We plotted antenna data for each spectral window (SPW) and again removed bad data. After removing the RFI and zero amplitude data, we checked for any spikes in the amplitude  and removed that bad data as well. The channels at the start and at the end were checked for RFI. We calibrated the data using the phase and flux density calibrators (Table~\ref{calibrator}). After each calibration, the solutions were checked by plotting them on {\tt plotcal}. Then we went back and removed the bad data and redid the calibrations. In general, SPW 0 and 7 in our data had problems. We did several rounds of calibration and plotted the corrected amplitude vs. phase of the flux density calibrators. When it showed a compact oval/circle on the amplitude-phase plane, we were satisfied with the calibration.

After satisfactory calibration, we imaged the data using the task CLEAN in CASA. 
Stokes I images of all 19 galaxies were produced with natural and
uniform weighting with different robust values at the respective center frequencies.
Spectral index ($\alpha$) maps were made using the two frequency images, which were first created using identical beams, pixel sizes and image sizes. The images were positionally aligned with the task 'OGEOM' and then combined with the task 'COMB' in AIPS. IMMATH task in CASA was also used to obtain the spectral indices. We used the IMFIT (CASA) and JMFIT (AIPS) tasks to fit Gaussians to the sources in order to obtain their flux densities and sizes (Tables \ref{single_core} and \ref{dual_core}).  Task IMEAN (AIPS) was also used to get the flux densities and the average values of spectral index from maps. 

\section{Results} \label{section5}

\subsection{Total Intensity and Spectral Index Images:} \label{section5.1}
We have imaged a total of 19 DPAGN with the VLA at different frequencies. We have found that three galaxies have dual radio cores, one galaxy shows kpc-scale core-jet structures and one shows extended structure which can be DAGN or core-jet structure of single AGN. Of the remaining 14 galaxies, 13 show single cores and one source is not detected. We have used the spectral index ($\alpha$) maps to understand the nature of these radio cores. Additionally, we have used the FIRST map to determine whether extended radio emission is present or not in the galaxies. For this purpose, we have calculated the ratio (Table \ref{gal_first}) of the integrated flux density to the peak intensity of FIRST map [$\theta_{FIRST}=(S_{int}/S_{peak})^{1/2}$] \citep{Singh2015} (Table \ref{gal_first}). The sources are defined to be extended if $\theta_{FIRST}\geq$ 1.06.
In this section, at first we describe our results for the galaxies that have clear dual radio cores and extended structures in our observations. Then we discuss all the single core galaxies together at the end of the section.

\subsubsection*{SDSS J100602.13+071130.9} J1006 (hereafter) is an interacting system. The projected nuclei separation is 5.2$^{\prime\prime}$ or 12 kpc.
The morphology classification of this galaxy and its companion is uncertain. 
Both nuclei have SDSS spectra. J1006 is classified as a star-forming, AGN broad line source. The companion has a QSO spectrum.
The radio images at 8.5 and 11.5 GHz (Figure \ref{J1006}) show two radio structures in this merger
galaxy system at a separation of $\sim$5$^{\prime\prime}$ or 12 kpc.
The radio emission of the main galaxy (A) is very faint ($\sim$0.06 mJy at 8.5 GHz). The upper limit in spectral index value is $\alpha$= $-0.93\pm$1.16.
The companion (B) shows an extended jet structure and the 
two hotspots are detected. The total size of the jet is 1$^{\prime\prime}$ or 2.4 kpc. 
The hotspots are in the NE-SW direction. One of the hotspots is brighter than the other which can be due to Doppler boosting.

\subsubsection*{SDSS J135558.08+001530.6} J1355 (hereafter) is a galaxy pair but the galaxy morphologies are uncertain. There is a single SDSS spectrum associated to this system which is classified as a
broadline AGN. We have detected two radio sources in the 8.5 and 11.5 GHz images (Figure \ref{J1355}) at a separation of $\sim$3.1$^{\prime\prime}$ or 8.06 kpc. Both the cores (A, B) are compact and the spectral index values are steep ($\sim$-1.0).

\subsubsection*{2MASX J16170895+2226279} J1617 (hereafter) is a spiral galaxy with a companion at 4.3$^{\prime\prime}$ (5.6 kpc) separation. Both galaxies have SDSS spectra. The J1617 spectrum has double-peaked emission lines but the companion does not have double-peaked emission lines. The optical image shows a common envelope in all bands and hence, this may be a minor merger. We have detected two cores at 6 GHz as well as in 15 GHz which coincide with the optical cores (Figure \ref{J1617}). The separation of the cores are 4.3$^{\prime\prime}$ or ~5.6 kpc. The primary core (A) has a spectral index value of $\alpha= -0.95\pm0.10$. This core has a one-sided jet in the 15 GHz image. The companion core (B) has a flat spectral index value $\alpha= -0.28\pm0.14$. 

\subsubsection*{SDSS J110215.68+290725.2} J1102 (hereafter) is an elliptical galaxy with a bright bulge and broad AGN emission lines. It has a two sided jet extended in the north (N) - south (S) direction. 
The total size of the core-jet structure is $\sim$ 2.9$^{\prime\prime}$ or 6.2 kpc in the 8.5 and 11.5 GHz images (Figure \ref{J1102}). The northern jet is composed of two blobs while the southern jet has a single blob. The southern blob is brighter than the northern one which can be due to Doppler boosting. However, the FIRST image of J1102 shows a large scale core-jet structure of size $\sim$ 120 kpc in the NE and SW directions (Figure \ref{J1102_FIRST}). This change in direction from the small scale radio jet to the large scale one is a signature of an Z-shaped core-jet structure in this galaxy (see section \ref{section6.2}). 
We overlaid the high-resolution radio image on the FIRST image (Figure \ref{J1102_FIRST}).
The spectral index map of the small scale radio structure shows that the core has a flat spectral index ($\alpha= 0.45\pm0.44$)  
and the jets have relatively steep indices ($\alpha< -1.81\pm0.46$). J1102 has been  observed in several surveys  
in the frequency range 0.365 to 4.85 GHz \citep[e.g.][]{Becker1991,Douglas1996,White1992}.

\subsubsection*{2MASX J14454130+3341080} J1445 (hereafter) is a QSO. 
J1445 shows a one-sided, small, extended structure at 8.5 GHz and 11.5 GHz (Figure \ref{J1445}). 
It contains a bright region with extended radio emission. The bright region has steep spectral index ($\alpha = -1.57\pm0.58$) and the extended region has $\alpha = -0.79\pm0.74$. The size of the radio emission is $\sim$ 0.44$^{\prime\prime}$ or 1.14 kpc. 
We have found a radio core in the FIRST image along with an extended low radio emission ($\sim 4\sigma$)
in the southwest direction. We are not sure whether this extended emission is a jet associated with the radio core or not.  
So, we have calculated the $\theta_{FIRST}$ which is $\approx 1.00$. This implies that most of the radio emission is concentrated 
at the center ($\sim 10$ kpc). Here, the extended radio emission at 8.5 and 11.5 GHz 
can be a core-jet structure or dual AGN (see section \ref{section6.2}).\\

We now discuss the remaining 14 sources (Table \ref{single_core}). The targets J1324 and J1349 have single frequency observations at 6 GHz. They both have compact cores. From their FIRST and NVSS images, we find that they
have similar flux densities at 1.4~GHz, which indicates that they are both compact sources. 

We have not been able to detect the source J1500 at any of the observed frequencies (8.5 and 11.5~GHz). However,
we have upper limits to the flux densities by taking 3 times the rms noise at 8.5 and 11.5 GHz (Table~5). We have made a
ball park estimate of the spectral index by using these flux densities. The calculated value is +0.44 which represents a radio source with an inverted spectral index. The VLA FIRST image shows an extended radio core. It has $\theta_{FIRST}$= 1.66 which also supports an extended jet emission. The SDSS spectrum indicates that it is an AGN. Hence, we can say that this is an AGN with low radio power which is not detected at the observed sensitivity.

\begin{figure*}
\includegraphics[width=0.74\columnwidth]{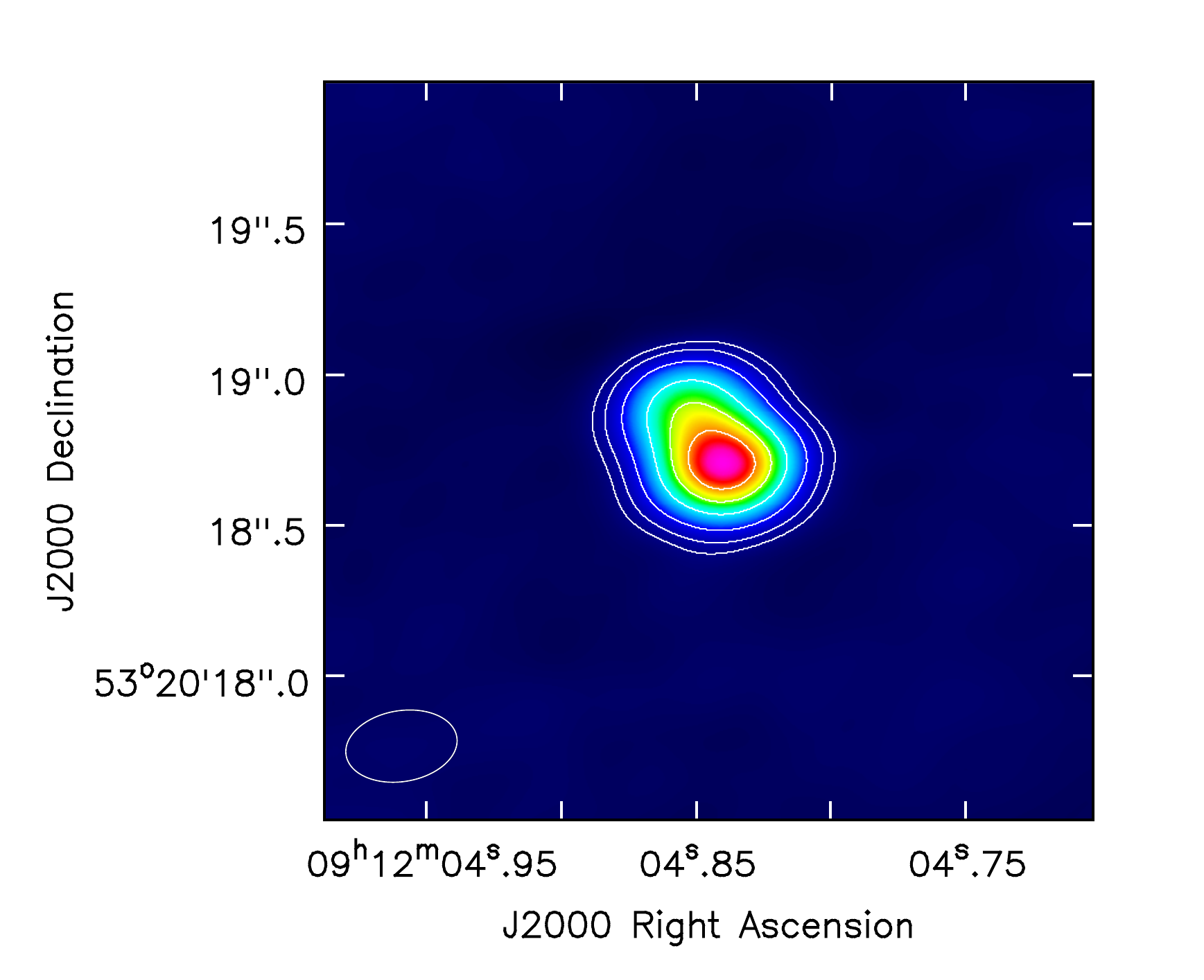}\includegraphics[width=0.71\columnwidth]{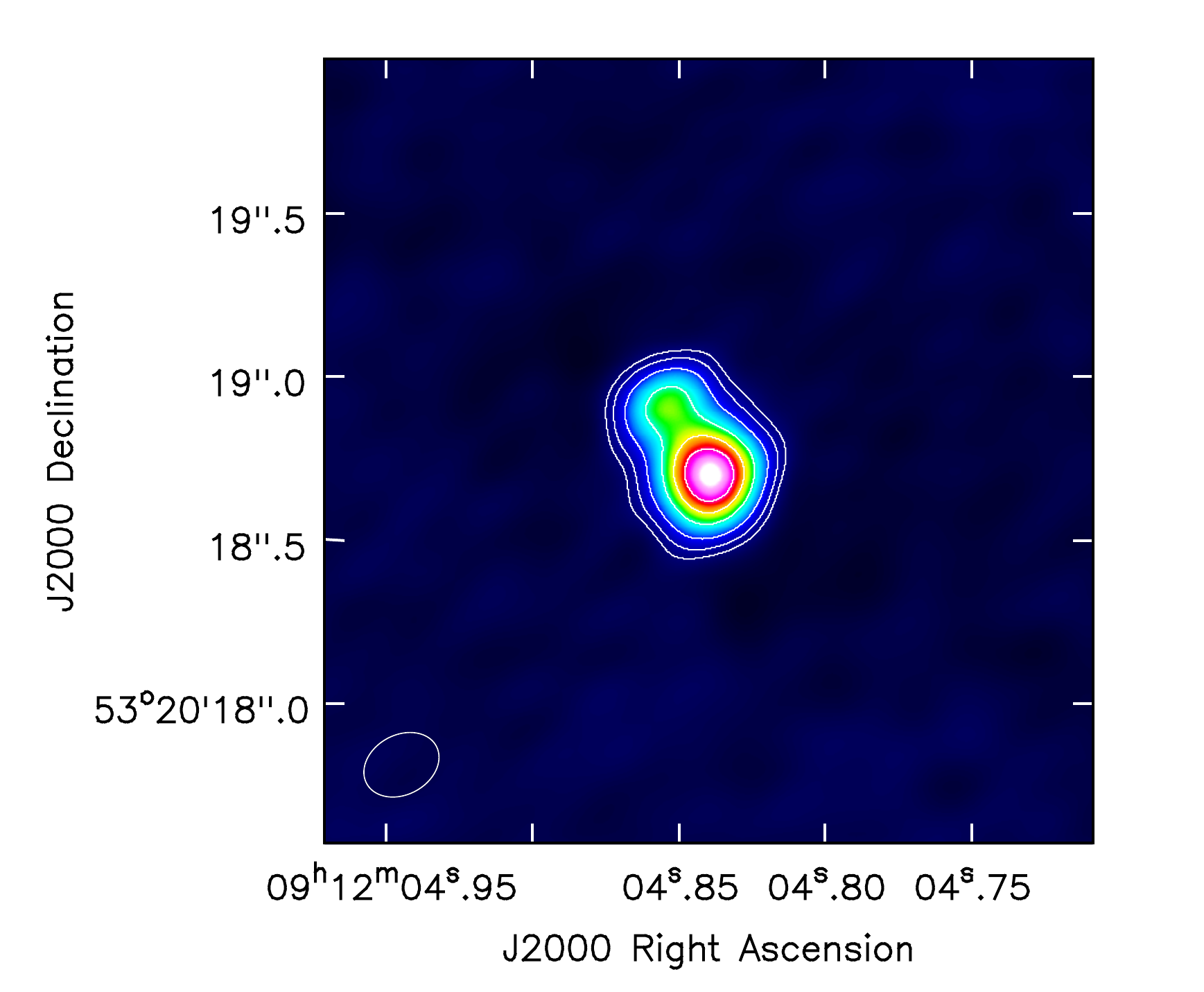}\includegraphics[width=0.7\columnwidth]{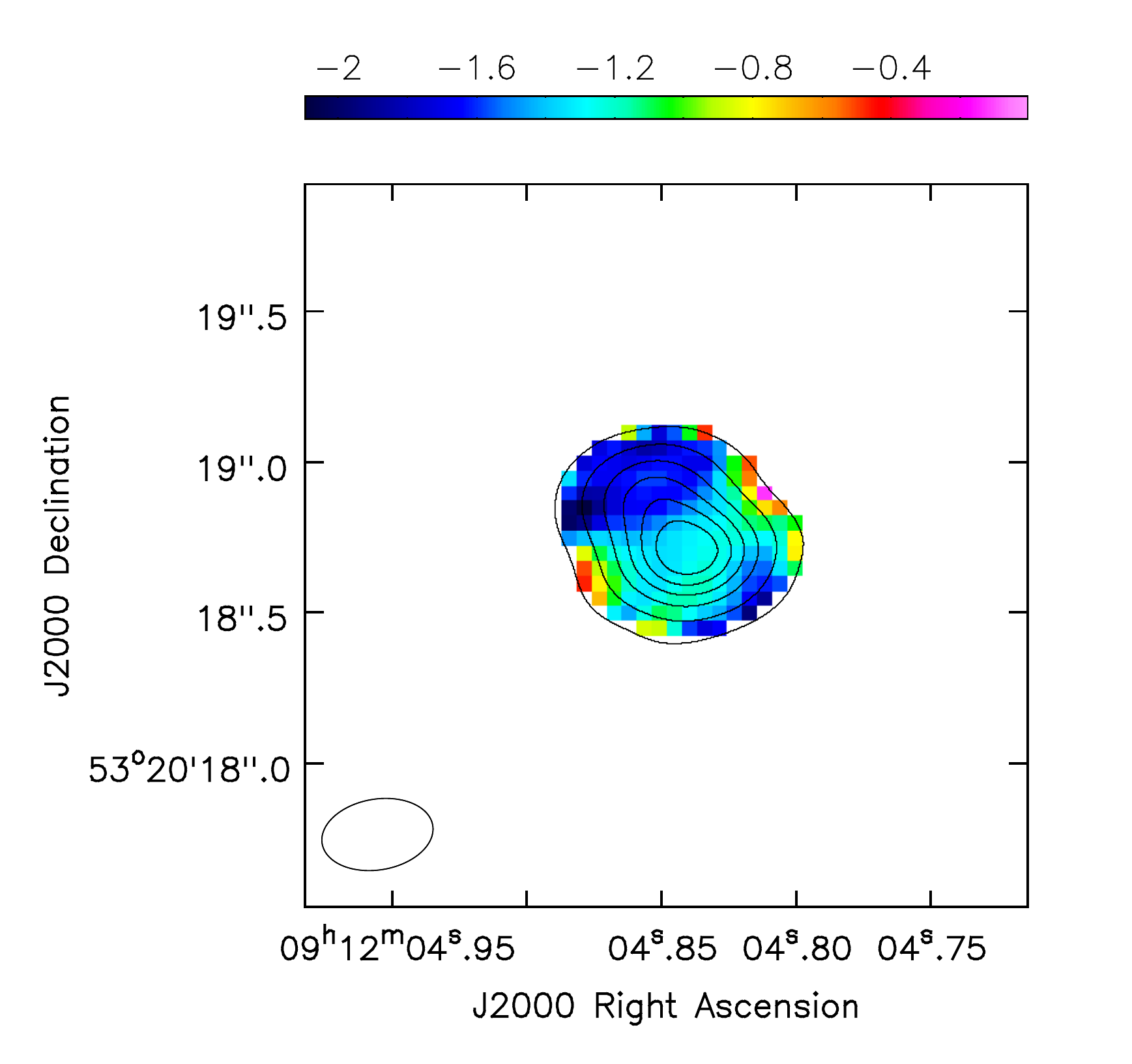}
\caption{The images of SDSS J091204.83+ 532018.5, situated 30$^{\prime\prime}$ away from our target source J0912.  Moving left to right is the 8.5 GHz image, 
11.5 GHz image and spectral index map using 8.5 GHz and 11.5 GHz images. The contour level are 10\%, 20\%, 40\%, 60\% and 80\% 
of its peak intensities respectively. We have resolved an extended radio structure in this object.}
\label{J0912_2nd}
\end{figure*}

We have detected flat spectral indices (Table \ref{single_core}) in six galaxies (J0912, J1413, J1420, J1504, J1600, J1644); indicative of compact AGN emission. 
However, in a recent VLBA study, \citet{Liu2018} have found that J0912 has a three component core jet structure on sub-arcsecond scales. There is an unresolved radio source at 30$^{\prime\prime}$ to the southeast of the J0912 in the FIRST image. This is the radio counterpart of the optical source SDSS J091204.83+ 532018.5 and its optical counterpart is very faint. We have detected an extended radio source with size of 0.75$^{\prime\prime}$ at the same position at 8.5 and 11.5~GHz (Figure \ref{J0912_2nd}). The bright core has a spectral index of $\alpha=-1.3\pm0.08$. The spectral index values increase towards the edges. The northeast region has $\alpha=-1.66\pm0.2$ and the southwest region has $\alpha=-1.46\pm0.4$. This suggests that this could be a kpc core-jet radio source (Figure \ref{J0912_2nd}). J1413 shows a large, two sided jet of size 275$^{\prime\prime}$ or $\sim$935 kpc in the NVSS image. J1504 has a jet of size 45~kpc in the FIRST image. The NVSS map of J1600 shows a jet of size 10 Mpc. J1420 and J1644 show compact cores in our observations and $\theta_{FIRST}\sim1.0$ which is also consistent with compact emission.

The remaining 5 sources have single steep-spectrum cores (UGC 05353, J1323, J1501, J2304 and J2336). UGC 05353, J1501 and J2336 do not have any extended jet emission in their FIRST or NVSS images. The calculated $\theta_{FIRST} < 1.06$ for UGC 05353 and J1501 also supports the lack of extended emission.
However, the steep spectral index indicates that optically-thin jet emission may be dominant in the cores. J2336 has $\theta_{FIRST} = 1.08$. Here again we cannot rule out extended jet emission inside the detected core. J1323 shows a core with steep spectral index value ($\alpha=-1.50\pm0.78$). J1323 shows extended radio emission in its NVSS image. Its FIRST image shows a two sided large-scale jet with a total size of ~550 kpc. We have detected a core and some radio emission in small blobs at the separation of $\sim8^{\prime\prime}$ (40 kpc) in the 8.5 and 11.5~GHz images of J1323. J2304 has a single core at the observed frequency (6, 8.5 and 11.5~GHz). \citet{Gabanyi2016} have observed this source with the Very Long Baseline Array (VLBA) at 1.5 GHz where they have detected a core-jet structure in the same direction as the 11.5 GHz image.

\begin{figure*}
 \includegraphics[width=2.0\columnwidth]{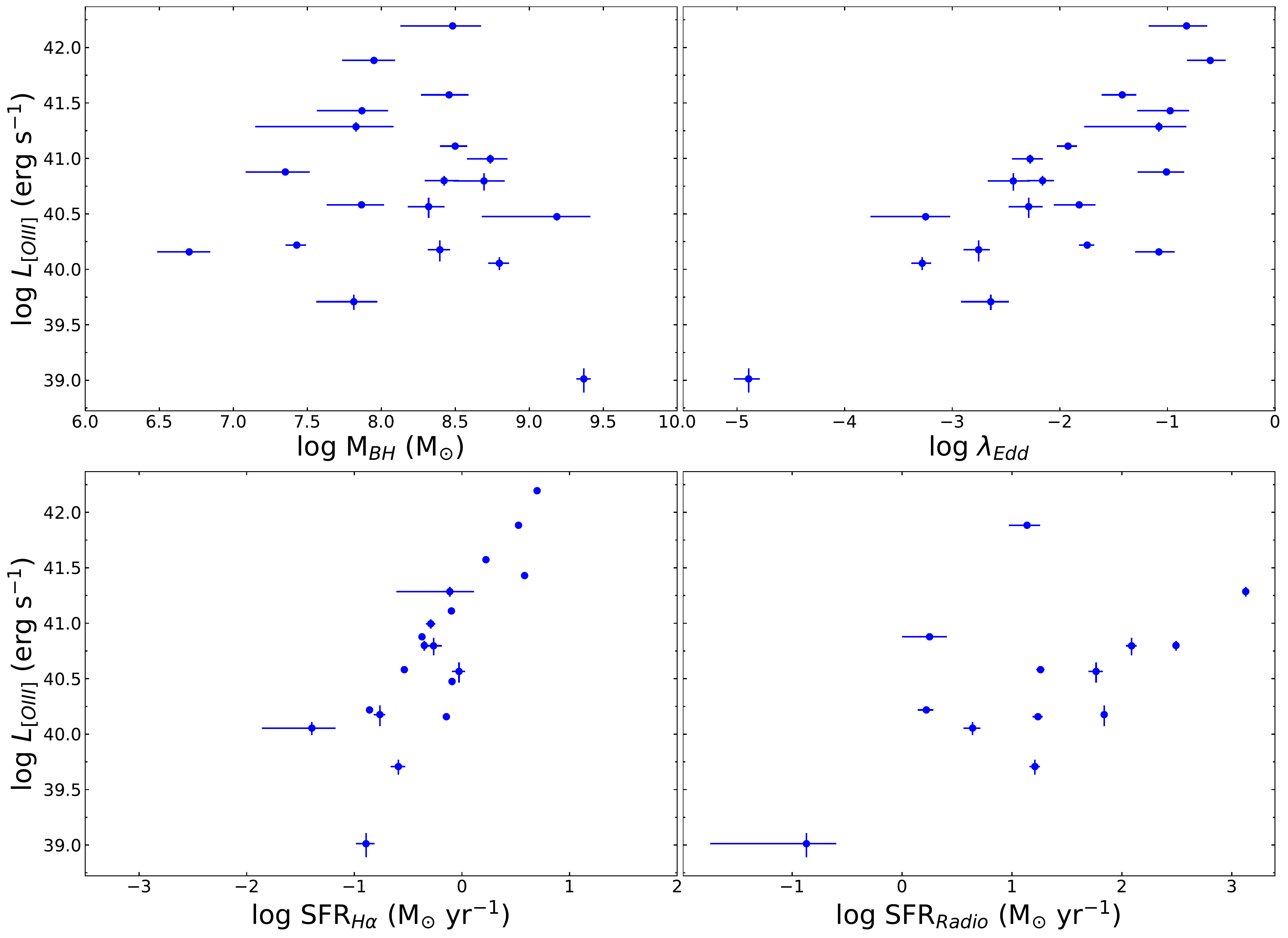}
\caption {The top left panel shows M$_{BH}$ vs L$_{[O III]}$. 
There is no correlation between the plotted quantities. Top right panel shows the $\lambda_{Edd}$ vs. L$_{[O III]}$. Here $\lambda_{Edd}$ falls in the range of single AGN $\lambda_{Edd}$ and there is a strong positive correlation with  L$_{[O III]}$. The bottom left panel shows the SFR$_{H\alpha}$ vs L$_{[O III]}$ which are also strongly correlated. The bottom right panel shows the SFR$_{Radio}$ vs L$_{[O III]}$. Errors have been plotted for all the quantities.}
\label{plot}
\end{figure*}

\subsection{SMBH masses, Eddington ratios and Star Formation Rates}\label{section5.2}
We have calculated the SMBH mass (M$_{BH}$) of all our sources. We have used the velocity dispersion ($\sigma$) of the underlying stellar component from \citet{ge.etal.2012} and the M-${\sigma}_\star$ relation from \citet{McConnell.etal.2013}. The SMBH values are given in (Table \ref{gal_prop}). The SMBH masses range from 10$^6$ to 10$^9$ M$_\odot$.  We calculated the total flux density using the blue shifted and red shifted component of the [O III]$\lambda$5008 line from \citet{ge.etal.2012}. Using source distances from NED\footnote{http://ned.ipac.caltech.edu/}, we obtained the [O III]$\lambda$5008 luminosities (L$_{[O III]}$). We used \citet{Heckman.etal.2004} to calculate the bolometric luminosities (L$_{Bol}$) of the sources. Using the SMBH masses, we calculated the Eddington luminosities (L$_{Edd}$) and the Eddington ratios ($\lambda_{Edd}$) (Table \ref{gal_prop}).
\begin{equation}
  L_{O[III]} = F_{[O III]} \times 4\pi D^2;\hspace{0.4cm} L_{Bol} = L_{[O III]} \times 3500\nonumber
\end{equation}
\begin{equation}
 L_{Edd} = 1.25 \times 10^{38} M_{BH} M_\odot^{-1};\hspace{0.4cm} \lambda_{Edd}=  L_{Bol}/L_{Edd}\nonumber
\end{equation}

We have used the $H\alpha$ flux from the SDSS spectra to calculate the star formation rate (SFR). The $H\alpha$ emission can arise from star formation as well as from the AGN. Therefore, the SFR calculated from the total $H\alpha$ is an upper limit. 
We have followed the SFR relation from \citet{Kennicutt.etal.1998}:
\begin{equation}
  SFR_{H\alpha} (M_\odot yr^{-1})=7.9\times10^{−42}L_{[H\alpha]}(erg s^{-1}) \nonumber
\end{equation}
Supernovae (Type II and Ib) remnants (SNRs) 
can give rise to accelerated relativistic electrons which produce non-thermal synchrotron emission. Massive stars in H II regions also produce free-free thermal emission. This radio emission results in the radio and far-infrared (FIR) correlation
observed for normal galaxies \citep{condon1992}. However, AGN host galaxies are offset from this correlation due to a ``radio excess''. We have used the radio excess objects which are AGN from
\citet{Yun2001} to obtain the expected AGN/SF fraction in radio flux density \citep[see Table 2 in][]{Yun2001} . To do this, we have calculated the expected radio emission using the FIR (F60 $\mu$m) flux density in the fitted correlation. Then we subtracted the calculated radio flux density from their observed values and found the AGN/SF contribution. We have done this exercise for all the objects in that table \citep[table 2;][]{Yun2001}. The AGN contribution in all the radio-loud objects is 99\% while in Seyfert galaxies, it is 80 to 90\%. Therefore, we have assumed the AGN contribution to have an upper limit of 70\% and the star formation contribution to be 30\%.

We have calculated the SFR using the radio flux density in the following relation \citep{condon1992}:
\begin{equation}
 SFR_{Radio} (M_\odot yr^{-1})= L(W Hz^{-1})/ 5.3\times10^{21} \nu(GHz)^\alpha \nonumber
 \end{equation}
We have used the radio flux density from the NVSS map at 1.4 GHz, assuming $\alpha=-0.8$ to calculate the SFR. Here, we have calculated the SFR$_{Radio}$ only for objects which have a single radio core. The SFR for the objects which show large core-jet structures are not calculated. This is because the SFR$_{Radio}$ would be much higher if the radio emission from the AGN jets are included. 

We have plotted the M$_{BH}$ vs L$_{[O III]}$, $\lambda_{Edd}$ vs L$_{[O III]}$, SFR$_{H\alpha}$ vs L$_{[O III]}$ as well as the SFR$_{Radio}$ vs L$_{[O III]}$ in Figure \ref{plot} and calculated the Pearson correlation (R) with probability (P) for each pair of variables. We find that M$_{BH}$ and L$_{[O III]}$ are not correlated and the correlation coefficient has a value R$=-0.11$, P$=0.64$. We find that the $\lambda_{Edd}$ follows the same trend as the single AGN control sample in \citet{Shangguan2016}.
It ranges from 0.013$\times10^{-3}$ to 0.249. The $\lambda_{Edd}$ values are correlated with L$_{[O III]}$, and has a value R= 0.79, P= $10^{-5}$. The SFR$_{H\alpha}$ is correlated with L$_{[O III]}$ and R= 0.82, P= $10^{-5}$. Although $\lambda_{Edd}$ is correlated with L$_{[O III]}$, we know that $\lambda_{Edd}$ depends on L$_{[O III]}$ and hence also with L$_{Bol}$. Therefore, we cannot say that this is a real correlation. However, the L$_{[O III]}$ and SFR$_{H\alpha}$ are independent parameters and hence this correlation is real. It may mean that star formation has been triggered by outflows or jet-ISM interaction \citep{Dugan2014}.

\section{Discussion}\label{section6}
We have detected dual radio cores in three of our sample double peaked emission line galaxies. In this section, we discuss the physical mechanisms that can explain both the nature of the optical spectra as well as the radio morphologies of the sources.

\subsection{DAGN in J1006, J1355 and J1617}\label{section6.1}
Observations of kpc scale DAGN can help us to constrain the theoretical merger models of galaxies \citep{Van2012,Blecha2013,Colpi2014}. \citet{Fu2015} have searched for kpc scale DAGN using a high resolution, 1.4~GHz VLA survey of the $\sim$92 deg2 SDSS
Stripe~82 field. They confirmed 4 DAGN in merger systems using both SDSS
spectra and VLA observations \citep{Fu2015L}. \citet{satyapal2017} used a
mid-IR pre-selection method to identify the DAGN candidates in merger systems
where the AGN are obscured by the presence of heavy dust. Since radio waves are
not obscured by dust, high resolution radio observations have helped to reveal many
DAGN \citep{mullersanchez2015,Fu2015}. However, the radio emission
can be due to nuclear star-burst as well as AGN activity. Hence, along with the
detection of two cores in radio images, the spectral index values of the
individual cores are needed to confirm the origin of the radio emission.
\citet{mullersanchez2015} have detected three DAGN with VLA where they have
found that the spectral indices of the cores are flat which is expected from the
AGN. On the other hand, \citet{Fu2015L} have found that the spectral indices of
the confirmed DAGN compact cores are steep. The steep spectral index found in compact steep spectrum (CSS) sources indicates that they are very young sources and probably have small scale jets.

The optical spectroscopy of the individual cores can also support the AGN scenario. The BPT diagram is one of the key tools to confirm or distinguish between the AGN or star-forming nature of the nuclei. The line ratios are used in the diagram to distinguish between the AGN and star-burst nuclei \citep{Baldwin1981}. However, in many cases, it is found that the emission lines are obscured by dust. In such cases, AGN can be detected with a mid-infrared color selection method using the {\em Wide-Field Infrared Survey Explorer (WISE)}\footnote{http://irsa.ipac.caltech.edu/Missions/wise.html} data which gives a higher detection rate compared to optical methods \citep{Satyapal2014}. In our study, we have used the spectral index values as well as the SDSS optical spectra (when available) to determine the nature of the dual cores in our sample galaxies. We have identified two cores in our images that we refer as the primary core or core A, which is generally identified with the optical position of the larger galaxy and the secondary core or core B which is usually associated with the companion galaxy.

Figure \ref{J1006} shows the 8.5 and 11.5 GHz images of J1006 which has two radio sources at the separation of $\sim$5$^{\prime\prime}$ or 12 kpc. As mentioned earlier, it is a merger system and both the sources are aligned with the optical cores (Figure \ref{J1006}). Source B shows a two sided jet with lobes/hotspots. The spectral index values of the lobes are steep ($\alpha$= $-1.3\pm0.4$ and $-1.8\pm0.9$). These spectral index values are expected from radio jets. We have calculated the line ratios, plotted them on the BPT diagram and found that it lies in the AGN region. Therefore, the radio images and optical spectra together confirm that source B is an AGN. The source A has $\alpha \leq-0.93\pm1.16$ and has a diameter of 0.37$^{\prime\prime}$ (0.72~kpc). The calculated radio power (log P$_{1.4}$) is $\sim$22 W Hz$^{-1}$ which is much lower than the expected radio power of young AGN (log P$_{1.4} \geq$ 25 W Hz$^{-1}$) \citep{ODea1998,An2012}. However, the optical emission line ratios in the BPT diagram support the AGN nature of this radio source \citep{ge.etal.2012}. Source A therefore, is a low radio power AGN. Thus, J1006 is a DAGN with an AGN separation of 12 kpc.

J1355 (Figure \ref{J1355}) is another galaxy merger where three optical nuclei can be clearly distinguished in the SDSS image of the galaxy. The optical sources are SDSS J135558.08+001530.6 (z= 0.134149), 2dFGRS N338Z121 (z= 0.1334) and SDSSCG 119 (z=0.1333). The separation of these sources from the main galaxy are 1.05$^{\prime\prime}$ and 5.87$^{\prime\prime}$. In our radio images, we have detected the main radio source at the position of SDSS J135558.08+001530.6 and the second radio source at a separation of 3.2$^{\prime\prime}$. However, it does not coincide with the other optical nuclei. Here we identify J1355, the radio source that coincides with an optical nuclei, as source A and the second radio core as source B. The spectral index value of source A is $-1.18\pm0.53$. The source is compact with size 0.4$^{\prime\prime}$. We have calculated the [O III]/H$\beta$ and [N II]/H${\alpha}$ and put them in the BPT diagram. It lies in the AGN region. So, from both the radio and optical data, we confirm that source A is an AGN. The spectral index of source B is $-0.97\pm1.07$. The size of the source is $\sim$1 kpc. This can be an AGN or a star-forming nucleus. The calculated radio power (log P$_{1.4}$) is $\sim$23 W Hz$^{-1}$. However, there is no optical source at the position of radio source B.  Hence, we cannot confirm the actual position of the source. It can be a background or foreground object. Without an optical spectrum of source B, we cannot rule out any of these origins. Hence, J1355 is a merger system containing either a DAGN or a AGN-SF nuclei pair at a separation of 3.1$^{\prime\prime}$ or 8.20 kpc provided the second radio source is at same redshift as the first source.

We have detected dual sources in J1617 at 6~GHz and 15 GHz (Figure \ref{J1617}) at a separation of 4.3$^{\prime\prime}$ or 5.6 kpc. This appears to be a minor merger. Here, we identify the main galaxy as source A and the accreting system as source B. The SDSS spectrum of source A shows double peaked emission lines with an AGN broad-line classification. We have made the spectral index image of source A which has an average $\alpha$ value of $-0.95\pm0.10$. We have detected a one-sided jet of size 1$^{\prime\prime}$ or 1.3 kpc. The steep spectral index value is consistent with jet emission. Therefore, we conclude that source A is a confirmed AGN. Source B has narrow emission lines in its SDSS spectrum and has a star-burst classification. The spectral index value of source B is $-0.28\pm0.14$, which is relatively flat and consistent with an AGN.  It is possible that the second SMBH is present but is dust obscured in the optical image. The calculated radio power (log P$_{1.4}$) is $\sim$22 W Hz$^{-1}$ which is much lower than expected AGN power \citep{An2012}. Hence, we cannot rule out a star-forming nucleus with present data. In previous studies of DAGN, SDSS J171544.05+600835.7 is an example of a system where in optical images only one nucleus is seen but with the help of spectroscopy and X-ray observations, this second nucleus has been found to host a DAGN and the AGN emission is obscured by dust \citep{Comerford2011}. \citet{Koss2016} have similarly discovered an obscured dual AGN in J2028.5+2543 using {\em Swift}, where both nuclei are heavily obscured to Compton-thick levels. Hence, to confirm the presence of a DAGN in J1355 and J1617, hard X-ray band observations are required.

\begin{figure}
 \includegraphics[width=.80\columnwidth,trim=30 0 160 20]{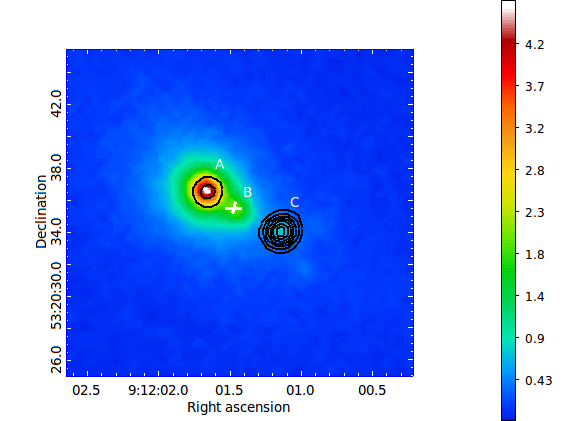}
\caption{ The optical g-band image of J0912 with the Chandra 0.7-3 kev X-ray image contours overlaid in black. The contour levels are 14\%, 28\%, 42\%, 57\%, 71\% and 85\% of the peak X-ray intensity. The optical image of J0912 shows three
cores (A, B, C) but we detected only one core (A) in our radio observations. In the
Chandra image, two cores (A and C) are detected at a separation of 10 kpc.}
\label{J0912_chandra}
\end{figure}

\subsection{Are J0192, J1445, J1102 dual AGN?} \label{section6.2}
\subsubsection*{2MASX J09120164+5320369}
The SDSS image shows that J0912 has 3 optical cores where
the secondary cores are at the separations of 2$^{\prime\prime}$ and 
5$^{\prime\prime}$  from the primary core ( A, B, C in Figure \ref{J0912_chandra}). We have detected a single
radio core in our 8.5 and 11.5 GHz images that coincide with the primary core (A)
in the optical image. There is Chandra archival data for this galaxy.
We found two X-ray cores in the Chandra image (Figure \ref{J0912_chandra}). One of the Chandra cores
coincides with the radio core associated with the primary optical
core (A). The second Chandra core coincides with a secondary optical core (C) at a
separation of 5$^{\prime\prime}$  or 10.165 kpc. The Chandra fluxes are $6.\times10^{-15}$ and  $3.96\times10^{-14}$ erg cm$^{-2}$ s$^{-1}$  respectively for the primary
and secondary cores \citep{Evans2010}. The redshift of this secondary core
is not available from NED or SDSS and we have not been able to found any spectra from this
core in the literature. If the secondary X-ray source is at the same redshift, then J0912 can be a DAGN at the separation of $\sim$ 10 kpc.

\subsubsection*{SDSS J110215.68+290725.2}
The 11.5~GHz radio image shows a core-jet structure of 2.8$^{\prime\prime}$
or 5.9 kpc for this source. Both sides of the jets are aligned. However, we have 
found that the FIRST (1.4 GHz) image shows a large scale core-jet structure
of size $\sim$60$^{\prime\prime}$ or $\sim$120~kpc. The large scale jets have a Z-shaped
structure (Figure \ref{J1102_FIRST}). There are many radio galaxies which show misaligned
jets \citep{Hutchings1988,Roberts2018}. Sometimes these mis-aligned
jets have S, Z or X shapes. The X-shaped sources in the literature \citep{Cheung2007}
have been explained by (i)~the spin-flip model \citep{Natarajan1998} or (ii)~the backflow model \citep{Leahy1984}. However, the S-shaped morphology is difficult to explain with these models. The S- or Z- shaped morphologies are better explained as helical or precessing jets \citep{begelman.etal.1980}. NGC~326 is a good example of a Z-shaped radio source \citep{Hodges.2012}, J1203 is another S-shaped galaxy \citep{Rubinur2017}. The precession of the jets in Z or S shaped sources can be due to (i)~the presence of a dual AGN system or (ii)~the tilted accretion disk of a single AGN.

\subsubsection*{2MASX J14454130+3341080}
J1445 shows an extended structure in which the bright component is likely to be the core and the extended emission is the radio jet. However, the spectral index image indicates the opposite to be true,
i.e. the bright region has a spectral index value of $-1.38\pm0.56$, whereas the extended region has
$\alpha=-0.79\pm0.74$. Therefore, it is possible that this system has two AGN where both have steep spectral index due to the core-jet structures. Alternatively it has a single core jet structure. We need high-resolution observations to distinguish between a DAGN or single core-jet structure in this radio source.

\subsection{Double-peaked emission lines in our dual radio source galaxies}\label{section6.3}
We have selected our sample based on their nuclear double-peaked emission lines expecting that these double-peaks may arise from the dual AGN. However, we have detected dual radio sources in 3 of our sample galaxies at the separation $\geq$ 3$^{\prime\prime}$; the SDSS fiber has a radius of 3$^{\prime\prime}$. Therefore, the dual radio structures cannot produce the double-peaks in the spectra of our sources as they lie at separations greater than the SDSS fiber diameter. Instead, jet-ISM interaction, outflows or NLR rotating disks can be the origin of the double-peaks. In the rotating disk model, the red-shifted and blue-shifted line components should have same flux density as they are from a single object \citep{Smith.etal.2010}. We cannot rule out the three possible origins of double-peaked lines only with radio data and SDSS spectra. \citet{mullersanchez2015} has mapped the [O III] region of their sample DPAGN galaxies. They have estimated the position angle of the [O III], position angle of the radio emission and the position angle of the galaxy. Due to lack of [O III] observation, we cannot rule out any one of the three possible origins of double-peaked
lines only with radio data and SDSS spectra.

Here, we discuss the three types of origin of the double-peaks in our detected dual radio source objects. J1006 has compact core (A). We have not detected any jet structure though the spectral index ($\alpha$) is steep. The SDSS spectra has a double peaked [O III] doublet. It has double-peak in H$\beta$. [O I] doublet are also present which indicate the presence of shock-heated gas \citep{Kharb2017a}. The [O III] double -peaks are not symmetric. Hence, it rules out their NLR disk. As radio jets were not detected, it is possible that outflows or shock heated gas can produce the double-peaks in the J1006 spectra. 

J1355 has a steep spectrum compact radio core (A). The [O III] double peaks are not symmetric. Multiple narrow line components are usually associated with outflows \citep{Crenshaw2010}. We find that the narrow lines in J1355 have multiple components. [O I]6363 is not present in the spectra. The AGN broad lines are present in the SDSS spectra. Hence, in J1355, outflows can give rise to the double-peaks.

J1617 has multiple peaks in the narrow lines. The [O~III] emission line double peaks are symmetric. \citet{ge.etal.2012} have fitted the spectra and found that the [O~III] has a blue-shifted wing with velocity 508 km~s$^{-1}$. We find that the radio image of J1617 shows a one sided jet at 15~GHz (A, Figure \ref{J1617}). Hence, the origin of the double-peaks can be either due to a rotating disk, outflows or jet-ISM interaction. None of these possibilities can be ruled out.

\subsection{The double-peaked emission lines in the remaining sources}\label{section6.4}

Two galaxies in our sample have extended radio emission. As we do not have [O III] maps of these sources, the direction of the [O III] regions cannot be determined. Hence, we cannot justify whether the jet is interacting with the [O III] region and producing the double-peaks. Therefore, jet-ISM interaction or outflows are the possible origin of double-peaks in these galaxies. At the same time,
if the double-peaks are symmetric then the NLR disk can also give rise to the double-peaks.
Multiple kinematic components in NLR, the presence of broad-lines or asymmetric wings are the the
signatures of outflows. For example, the radio source J2304 has been studied earlier using the VLBA \citep{Gabanyi2016}. They found that the DPAGN lines in J2304 can arise from jet- ISM interaction as they
did not detect a second AGN or radio core. Furthermore,  AGN broad- lines are also present
in the SDSS spectrum of J2304. We found that all of our sample galaxies have broad emission lines present in their SDSS DR12 \footnote{www.sdss.org/dr12/} spectra. So, there is a possibility that AGN outflows are creating the double-peaks
in this source as well as the remaining sources in our sample. 

\subsection{The detection of dual AGN in our sample}\label{section6.5}
We have observed 19 DPAGN, expecting that the double peaks are due to dual AGN and we would be able to resolve those DAGN with high resolution radio observation. We have detected one confirmed DAGN but this DAGN is not responsible for the double peaks in the sample galaxy. As discussed in the section \ref{section6.3}, the detected DAGN is at the separation $\geq~3^{\prime\prime}$ which is the size of the SDSS optical fiber. The detected DAGN is in a merging galaxy systems. Hence, we can say that the detection of DAGN from our DPAGN sample is zero and the detection is in merging galaxies only. 

\citet{Shen2011} have found that 10\% of the DPAGN in their sample are DAGN. \citet{mullersanchez2015} have observed their sample at 8.5~GHz and 11.5~GHz. They have detected 3 dual AGN which is 15\% of their sample. About 75\% of the DPAGN in their sample arise from jet-ISM interactions. They have selected their sample from long-slit observations of DPAGN. \citet{Tingay2011,Gabanyi2016,Liu2017} have conducted VLBA studies of DPAGN sources but no binary/dual AGN were detected. \citet{Kharb15,Kharb2017a,Kharb19} have studied the DPAGN KISSR~1494, KISSR~1219 and KISSR~434 with the VLBA and found that the double-peaks are likely due to outflow-ISM interaction. Recently, \citet{Fu2015L} have searched for DAGN using the SDSS and FIRST images of the Stripe 82 degree region on the sky. They have searched for close companions or merger systems. They have observed that reliable dual core galaxies have SDSS AGN spectra that are associated with the individual cores in the VLA images. 

\citet{Fu2015L} have detected 4 dual AGN out of the 6 merging galaxies. From our observations also, we have found that most of the DPAGN are due to the jet-ISM kinematics or outflows. Hence, DPAGN are not good tracer of DAGN. \citet{satyapal2017} have also pointed out that many AGN are obscured in optical due the heavy dust. Hence, selecting the sample using SDSS low resolution optical spectra is not a good idea. They have used the IR color for their sample selection of DAGN and they have subsequently carried out Chandra observations of the sample. The detection rate is quite high in their sample. Although it is likely that the sample of merging galaxies would not always provide very close DAGN ($\leq$3$^{\prime\prime}$) but there is a high probability of DAGN detection at the separation $\leq$ 10 kpc.

\section{Summary}\label{section7}
\begin{enumerate}
\item We have selected 20 DPAGN galaxies on the basis of $\Delta V$ vs $\sigma$ plot to search for dual AGN. We have studied these galaxies using high resolution VLA observations. The source 2MASXJ 12032061+1319316 shows a clear S-shaped radio jet suggesting precession \citep{Rubinur2017}. The remaining 19 sources are discussed in the present paper.

\item From the radio morphologies, we find that three galaxies in our sample of 19 galaxies are dual radio sources at the separations of $\lesssim$ 10kpc, other two have extended radio structures. The remaining 13 galaxies have single radio sources and one galaxy is not detected.

\item The dual radio source galaxies are SDSSJ100602.13+071130.9, SDSSJ135558.08+001530.6 and 2MASXJ16170895+2226279. The radio sources have projected separations of 12, 8.2 and 5.6 kpc, respectively. All three galaxies are in merger systems.

\item From the spectral index map and SDSS spectra we have confirmed that radio sources in J100602.13+071130.9 form a dual AGN system. The primary object is a radio-weak AGN while the secondary has resolved two sided jets in 8.5 and 11.5 GHz images. 

\item SDSSJ135558.08+001530.6 has two steep spectrum cores which can be a dual AGN or an AGN -starforming nuclei pair.

\item  2MASXJ16170895+2226279 has two cores where the primary core has a steep spectrum which is because of jet contribution detected at 15 GHz. The second core has a flat spectral index. This also can be an AGN or star-forming nucleus.

\item SDSSJ110215.68+290725.2 and 2MASXJ14454130+3341080 have extended radio structures in our observations at 8.5 and 11.5 GHz. SDSS J110215.68+290725.2 has a two-sided core-jet structure of size $\sim$6~kpc. This source has a large-scale jet of size 132 kpc in its VLA FIRST image. We found a directional change in the radio jet from the small scale (few kpc) to the large scale (100 kpc) image and it appears to have a Z shape. This radio structure can be due to precession of the jet caused by a dual/binary AGN system or a tilted accretion disk associated with a single AGN. The source 2MASXJ14454130+3341080 has a bright core and is extended out on one side to a radius of 1.14~kpc. Both the core and the extended structure have steep spectral indices. 

\item We determined the SMBH masses, Eddington ratios and the star formation rates using the optical spectra and the radio emission (1.4~GHz). We found that the [O III] luminosity (L[O~III]) is strongly correlated with the SFR(H$\alpha$). 

\item As our detected dual AGN is at a separation of $>3^{\prime\prime}$, we believe that the DPAGN in our sample do not originate from the dual AGN in these systems. Jet-ISM interaction, outflows or NLR rotating disk give rise to the DPAGN in these galaxies.

\item High resolution radio observations alone cannot help us to determine the origin of the DPAGN in our sample galaxies. We deduce that most of the DPAGN arise from NLR-jet kinematics as we have not detected dual AGN in these systems. However, we can not rule out the possibilities of a faint second AGN or a nearby AGN which is below our resolution and sensitivity limits.

\item We find that DPAGN in low resolution optical spectra such as those in SDSS, are not good indicators of dual/binary AGN. Instead, closely interacting galaxies merger remnants are good candidates for detecting dual/binary AGN.
\end{enumerate}

\section{ACKNOWLEDGEMENTS}
We thank both the referees for their insightful comments that have significantly improved this paper. We acknowledge IIA for providing the computational facilities.
RK wants to thank Dr. Sumana Nandi, Avinash Singh and Prasanta Kumar Nayak for helpful discussions.
The National Radio Astronomy Observatory is a facility
of the National Science Foundation operated under cooperative
agreement by Associated Universities, Inc.
This research has made use of the NASA/IPAC Extragalactic Database (NED),
which is operated by the Jet Propulsion Laboratory, California
Institute of Technology, under contract with the National Aeronautics and Space Administration.
Funding for the Sloan Digital Sky Survey IV has been provided by 
the Alfred P. Sloan Foundation, the U.S. Department of Energy Office of Science,
and the Participating Institutions. SDSS- IV acknowledges support and resources 
from the Center for High-Performance Computing at the University of Utah. 
The SDSS web site is www.sdss.org.
We acknowledge the usage of the HyperLeda database (http://leda.univ-lyon1.fr)

 \bibliographystyle{mn2e}

\begin{table*}
\centering
\caption{{\bf List of the confirmed DAGN with SDSS spectra from literatures (red points in Figure \ref{delv_sig}):} Column 1: SDSS name of the galaxy; Column 2: Velocity difference between 
the red-shifted and blue-shifted [O III] emission line ($\Delta V$); Column 3: The velocity dispersion of 
the stellar population in the galaxies ($\sigma$); Column 4: References corresponding to the detection of the DAGN; **List of the DPAGN which could not yield DAGN after higher resolution observations (blue points in Figure \ref{delv_sig}) are added at the end. {\bf Note:} SDSS J1023+3243 is a confirmed DAGN based on VLA data by M$\ddot{u}$ller-S$\acute{a}$nchez et al. (2015), however, subsequent VLBA imaging by Gab$\acute{a}$nyi et al. (2016) failed to detect any really compact radio feature in this object.}
\label{conf_dagn}
\resizebox{0.97\textwidth}{!}{
\begin{tabular}{ccccccc}
\hline Name        &  \begin{tabular}[c]{@{}c@{}}Velocity difference\\ (km~s$^{-1}$)\end{tabular} &  \begin{tabular}[c]{@{}c@{}}Velocity Dispersion\\ (km~s$^{-1}$)\end{tabular}  & Reference\\ \hline
SDSS J0952+2552 & 599  & 273 & \citet{McGurk.etal.2011} \\
SDSS J1023+3243 & 371  & 281  & \citet{mullersanchez2015} \\
SDSS J1108+0659 & 25   & 359  &  \citet{Liu2013} \\
SDSS J1126+2944 & 191  & 310    &  \citet{Comerford2015} \\
SDSS J1146+5110 & 156  & 282  &  \citet{Liu2013}\\
SDSS J1158+3231 & 372    & 441    &  \citet{mullersanchez2015}\\
SDSS J1332+0606 & 355  & 429   &  \citet{Liu2010} \\
SDSS J1502+1115 & 400    & 647   & \citet{Fu2011}   \\
SDSS J1623+0808 & 400    & 469 &  \citet{mullersanchez2015}\\
SDSS J1715+6008 & 197  & 358    & \citet{Comerford2011} \\ \hline\hline
**Non-confirmed   & DPAGN   & with & higher resolution obs\\ \hline \hline
SDSS J0002+0045 & 486  &221 &\citet{mullersanchez2015} \\
SDSS J0731+4528 & 276 &155 & \citet{mullersanchez2015}\\
SDSS J0736+4759 & 257 &175 &\citet{mullersanchez2015} \\
SDSS J0802+3046 & 349 & 158 &\citet{mullersanchez2015} \\
SDSS J0846+4258 &510  &213 &\citet{mullersanchez2015} \\
SDSS J0858+1041 & 257 & 207& \citet{mullersanchez2015}\\
SDSS J0930+3430 & 257& 144 & \citet{mullersanchez2015}\\
SDSS J1023+3243 & 355 &155 &\citet{mullersanchez2015}\\
SDSS J1027+3059 & 207 & 168  &\citet{mullersanchez2015} \\
SDSS J1112+2750 &335 & 212 & \citet{mullersanchez2015}\\
SDSS J1152+1903 &  236 & 146&\citet{mullersanchez2015} \\
SDSS J1556+0948 & 391 & 201 &\citet{mullersanchez2015} \\
SDSS J1623+0808 & 426&  181& \citet{mullersanchez2015}\\
SDSS J2254-0051 &352 & 188 &\citet{mullersanchez2015} \\
SDSS J0854+5026 & 320  & 130 &   \citet{Comerford2015}\\
SDSS J1006+4647 &  229 & 171 &  \citet{Comerford2015}\\
SDSS J1126+2944 & 309 & 169 & \citet{Comerford2015}\\
SDSS J1322+2631 & 397 & 184 & \citet{Comerford2015}\\
SDSS J1604+5009 & 368 & 189 & \citet{Comerford2015}\\
SDSS J0752+2736 & 281 & 119 & \citet{Comerford2015}\\
SDSS J1356+1026 &413 & 238 & \citet{Comerford2015}\\
SDSS J1448+1825 & 251 & 139 & \citet{Comerford2015}\\
SDSS J1137+6120 &      237  & 301  & \citet{Liu2017}\\  
SDSS J1243-0058 &      521  & 311 &  \citet{Liu2017}  \\
SDSS J1352+6541 &      373  & 268 &  \citet{Liu2017}   \\
SDSS J2310−0900 &   327   & 187 & \citet{Liu2017}\\      
SDSS J2333-0049  &        516 &     90 &\citet{Liu2017}\\
SDSS J0009-0036   &     332   &   203 & \citet{Liu2017}\\
SDSS J0738+3156    &    297   &   258 & \citet{Liu2017} \\
SDSS J0803+3926   &      391   &   192&  \citet{Liu2017}\\
SDSS J0858+1041   &     384    &   222 &\citet{Liu2017}\\
SDSS J1356+1026   &     413   &    238& \citet{Liu2017}\\
SDSS J1715+6008   &    347    &  131 & \citet{Liu2017}\\ \hline\hline

\end{tabular}
}
\end{table*}

\begin{table*}
\centering
\caption{{\bf Properties of our sample galaxies:} Column 2: Name of the sample sources; Column 3,4: J2000 RA and DEC coordinates; 
Column 5: Redshifts of the sources; Column 6: Scale at the Hubble flow distance; 
Column 7: Luminosity distance; Column 8: Velocity difference between 
the red-shifted and blue-shifted O[III] emission line ($\Delta V$); Column 9: The velocity dispersion of 
the stellar population in the galaxies ($\sigma$).}
\label{sam_gal}
\resizebox{\textwidth}{!}{
\begin{tabular}{ccccccccccccc}
\hline\begin{tabular}[c]{@{}c@{}}Object\\ No\end{tabular} & Name                     & \begin{tabular}[c]{@{}c@{}}Right   \\ Ascension\end{tabular} & Declination  & z    & {\bf Scale} (kpc/${\prime\prime}$)    & \begin{tabular}[c]{@{}c@{}}Distance \\ {\bf (Mpc)}\end{tabular}  & \begin{tabular}[c]{@{}c@{}}$\Delta V$\\(km~s$^{-1}$)\end{tabular} &\begin{tabular}[c]{@{}c@{}}$\sigma$\\(km~s$^{-1}$)\end{tabular} \\\hline\hline
1                                                  & UGC 05353                   & 09:58:40.09                                                 & +28:52:39.22 & 0.021     &    0.417                  &   92                                           & 492.1              &  306.9                             \\
2                                                   & 2MASX J13245059+1758152    & 13:24:50.59                                                 & +17:58:15.04 & 0.079     &    1.711                     &   381                                              & 534.6              &   103.3                           \\
3                                                   & 2MASX J13490964+0404487    & 13:49:09.64                                                   & +04:04:48.87    & 0.085     &     1.582                   &   351                                                & 445.3             &   206.3                            \\
4                                                   & 2MASX J16170895+2226279    & 16:17:08.95                                                   & +22:26:27.00  & 0.065     &      1.313                &   284                                              & 450.6              &   134.7                          \\
5                                                   & 2MASX J16441390+2528286    & 16:44:13.90                                                   & +25:28:28.60    & 0.055     &      1.113               &   238                                             & 463.7              &   243.1	                          \\
6                                                  & 2MASX J23044283-0933454    & 23:04:42.83                                                   & -09:33:45.40    & 0.032     &       0.645                &   130                                            & 307.8              &   139.0                          \\ 
7                                                   & SDSS J233604.04+000447.1  & 23:36:04.04                                                     & +00:04:47.10    & 0.076     &        1.532                 &   327                                             & 472.0             &   166.2                    \\   \hline                              
8                                                   & 2MASX J09120164+5320369    & 09:12:01.68                                                  & +53:20:36.90  & 0.101     &       2.033                 &   454  &  418.4  & 208.7\\
9 & SDSS J100602.13+071130.9                                                   & 10:06:02.13 & +07:11:30.90  &   0.121            &     2.407                   &   550         & 596.7  &  166.4\\
10    & SDSS J110215.68+290725.2                                                  & 11:02:15.68 & +29:07:25.24    & 0.106           &    2.111              & 476    & 417.2  &  215.2\\
11    & SDSS J132318.81+030807.1                                                  & 13:23:18.82 & +03:08:07.13   & 0.269            &     5.353             &   1327  & 465.2 & 213.7\\
12    & SDSS J135558.08+001530.6                                                  &  13:55:58.09 & +00:15:30.60   & 0.134           &     2.668                 & 612  & 531.6  & 284.9\\
13    & 2MASX J14131625+2119374                                                   & 14:13:16.25  & +21:19:37.47   &  0.172         &     3.444               & 806 & 427.7  &  237.1 \\
14    & 2MASX J14203147+4008166                                                  & 14:20:31.51  & +40:08:15.97    & 0.210           &      4.207              & 1005  & 461.0  &  233.0\\
15    & 2MASX J14454130+3341080                                                   &  14:45:41.30 & +33:41:07.86    & 0.131           &       2.615                 &  595   & 494.5 & 172.0\\
16    & 2MASX J15001769+1051100                                                   &   15:00:17.73 &  +10:51:09.81  &  0.170          &      3.404                 & 795   & 443.5 &  200.0 \\
17    & B31459+399NED01                                                          &   15:01:02.57  &   +39:42:00.07  &  0.355        &       7.077               & 1828  & 565.5 & 163.7\\
18    & 2MASX J15042218+4741116                                                   & 15:04:22.21   & +47:41:12.06     & 0.093         &        1.868                    & 413  &  418.5  & 260.9 \\
19    & SDSS J160027.78+083743.0                                                   & 16:00:27.78   & +08:37:43.00    & 0.226         &         4.518                 & 1089    & 418.2  & 211.6\\ \hline \hline
\end{tabular}
}
\end{table*}

\begin{table*}
\centering
\caption{{\bf Observation catalogue:}}
\label{obs_cat}
\resizebox{\textwidth}{!}{
\begin{tabular}{cccccccc}
\hline Observation ID & \begin{tabular}[c]{@{}c@{}}Observed Frequency\\(GHz)\end{tabular} 
& Configuration & \begin{tabular}[c]{@{}c@{}}Expected resolution \\($^{\prime\prime}$)\end{tabular}&  \begin{tabular}[c]{@{}c@{}}Observed time\\(h) \end{tabular}  \\ \hline
15A-068        &   6.0             & A             &0.33  & 2.00 \\ 
16A-144        &   15.0              & B             &0.42    &  2.00  \\
16B-002        &   8.5 and 11.5    & A             &0.20    & 2.67 \\
13B-020 (NRAO archive)    &8.5 and 11.5      & A               & 0.20   & --\\ \hline

\end{tabular}
}
\end{table*}

\begin{table*}
\centering
\caption{{\bf List of flux density and phase calibrators in the observations:} Column 1: SDSS name of the galaxy; Column 2: VLA project ID; Column 3: Name of the flux density calibrator; Column 4: Name of the phase calibrator; Column 5: Date of observation.}
\label{calibrator}
\resizebox{\textwidth}{!}{
\begin{tabular}{ccccccc}
\hline Name    & Project ID  &  \begin{tabular}[c]{@{}c@{}}Flux density calibrator \end{tabular} &  \begin{tabular}[c]{@{}c@{}}Phase calibrator\end{tabular}  &DOB \\ \hline

2MASXJ09120164+5320369  & 16B-002  & 3C147  &J0854+5757 &15th Nov 2016     \\
UGC 05353               & 15A-068  & 3C147  & J0956+2515   &19th July 2015 \\
                        & 16A-144  & 3C286  & J1013+2449  & 20th May 2016    \\
SDSS J100602.13+071130.9  & 16B-002  &3C147 &J0954+1743   & 15th Nov 2016     \\             
SDSS J110215.68+290725.2  &  16B-002 &3C147 & J1125+2610  & 15th Nov 2016      \\
SDSS J132318.81+030807.1  &  16B-002 & 3C286 & J1354-0206 & 7th Jan 2017           \\
2MASX J13245059+1758152   &  15A-068 & 3C147 & J1327+2210 & 19th July 2015     \\
2MASX J13490964+0404487   &  15A-068 & 3C147 &  J1354-0206& 19th July 2015  \\
SDSS J135558.08+001530.6  &  16B-002 &3C286  &J1354-0206  & 7th Jan 2017    \\
2MASX J14131625+2119374   & 16B-002  & 3C286 & J1436+2321 & 7th Jan 2017   \\
2MASX J14203147+4008166   & 16B-002 & 3C286 & J1416+3444  & 7th Jan 2017      \\
2MASXJ14454130+3341080    &  16B-002 & 3C286 & J1416+3444  & 7th Jan 2017      \\
2MASX J15001769+1051100   &  16B-002 & 3C286  & J1504+1029 & 7th Jan 2017  \\
B31459+399NED01           & 16B-002  & 3C286 & J1500+4751 &  7th Jan 2017  \\
2MASX J15042218+4741116   & 16B-002  & 3C286 & J1500+4751 &  7th Jan 2017  \\  
SDSS J160027+083742       &  16B-002 &3C286 & J1608+1029  & 7th Jan 2017    \\
2MASX J16170895+2226279   &  15A-068 & 3C147 & J1613+3412 & 19th July 2015 \\
                          & 16A-144  & 3C286 & J1613+3412 & 29th May 2016 \\
2MASX J16441390+2528286   &  15A-068 & 3C147 & J1613+3412 & 19th July 2015   \\
                          & 16A-144   & 3C286 & J1613+3412 & 29th May 2016  \\
2MASX J23044283-0933454   &  15A-058  & 3C48 & J0137+3309 & 25th June 2015 \\
                          & 13B-020  & 3C48 & J2323-0317 & 20th Feb 2015  \\
SDSS J233604.04+000447.1  &  15A-058 &  3C48 & J2323-0317 & 25th June 2015 \\
                          &16A-144 &  3C48& J2320+0513 & 20th May 2016  \\\hline

\end{tabular}
}
\end{table*}

\begin{table*}
\centering
\caption{{\bf Radio properties of the single AGN in our sample:} Column 1: Name of the galaxies; column 2: the observation frequencies;
column 3: The beamsize: Major axis ($\theta_{1}$), minor axis ($\theta_{2}$) and position angle (PA);
column 4: The RMS noise
 in $\mu$Jy/beam; column 5: the flux density (S$_{int}$) in the intensity images in mJy;
 column 6: the spectral index value ($\alpha$); 
{\bf Note:} (a)  An upper limit of $\alpha$ is given for extended sources. (b) $\alpha$ of the J1102 core is given.}
\label{single_core}
\resizebox{\textwidth}{!}{
\begin{tabular}{cccccccccccccccc}
\hline\hline Name    & \begin{tabular}[c]{@{}c@{}}Frequency\\ (GHz)\end{tabular}    &\begin{tabular}[c]{@{}c@{}} Beamsize \\($\theta_{1} \times \theta_{2}$, PA)  \end{tabular}  & \begin{tabular}[c]{@{}c@{}}Noise\\ ($\mu$Jy/beam)\end{tabular} & \begin{tabular}[c]{@{}c@{}}S$_{int}$\\(mJy)\end{tabular} & \begin{tabular}[c]{@{}c@{}}$\alpha$\end{tabular}  \\\hline\hline

2MASX J09120164+5320369  & 8.5        &0.37$^{\prime\prime}\times$0.24$^{\prime\prime}$, -80.62$^\circ$    & 15.79 & 62.80 & -0.43$\pm$0.01 \\ 
		         & 11.5       &0.24$^{\prime\prime}\times$0.18$^{\prime\prime}$, -61.63$^\circ$    & 16.27  & 55.50 &\\ \hline
UGC 05353                & 6.0        &0.43$^{\prime\prime}\times$0.29$^{\prime\prime}$, 76.55$^\circ$     & 13.02 & 3.82  &-0.82$\pm$0.03 \\
                         & 15.0       &0.52$^{\prime\prime}\times$0.38$^{\prime\prime}$, 78.14$^\circ$     & 7.76  & 1.79  &  \\ \hline
SDSS J110215.68+290725.2 & 8.5        &0.24$^{\prime\prime}\times$0.19$^{\prime\prime}$, -67.32$^\circ$    & 15.42 & 2.40  &$\leq$ -1.81$\pm0.46^{a}$  \\
                         & 11.5       &0.17$^{\prime\prime}\times$0.15$^{\prime\prime}$, -19.23$^\circ$    & 15.99 & 1.64  & 0.45$\pm0.44^{b}$ \\ \hline
SDSS J132318.81+030807.1 & 8.5        &0.38$^{\prime\prime}\times$0.21$^{\prime\prime}$, 87.56$^\circ$     & 19.67 & 1.57  &-1.50$\pm$0.78 \\ 
			 & 11.5       &0.28$^{\prime\prime}\times$0.17$^{\prime\prime}$, 88.53$^\circ$     & 20.70 & 1.01  &   \\ \hline
2MASX J13245969+1758152  & 6.0        &0.32$^{\prime\prime}\times$0.30$^{\prime\prime}$, -59.36$^\circ$    & 13.20 & 4.06  & -  \\ \hline 
2MASX J13490964+0404487  & 6.0        &0.35$^{\prime\prime}\times$0.31$^{\prime\prime}$, -38.88$^\circ$    & 13.67 & 19.39 & - \\ \hline
2MASX J14131625+2119374	 & 8.5        &0.25$^{\prime\prime}\times$0.19$^{\prime\prime}$, 84.45$^\circ$     & 24.80 & 2.11  &-0.31$\pm$0.18  \\
                         & 11.5       &0.19$^{\prime\prime}\times$0.15$^{\prime\prime}$, 83.12$^\circ$     & 28.50 & 1.91  & \\ \hline
2MASX J14203147+4008166  & 8.5        &0.24$^{\prime\prime}\times$0.16$^{\prime\prime}$, -64.74$^\circ$    & 22.13 & 3.33 & -0.23$\pm$0.20 \\                  
                         & 11.5       &0.19$^{\prime\prime}\times$0.12$^{\prime\prime}$, -62.17$^\circ$    & 25.30 & 3.11  & \\ \hline
2MASX J14454130+3341080  & 8.5        &0.22$^{\prime\prime}\times$0.18$^{\prime\prime}$, -60.66$^\circ$    & 21.35 &  0.62 & $\leq$-1.40$\pm1.16^{a}$  \\  
                         & 11.5       &0.17$^{\prime\prime}\times$0.14$^{\prime\prime}$, -59.20$^\circ$    & 24.36 & 0.40   & \\ \hline
2MASX J15001769+1051100  & 8.5        &0.30$^{\prime\prime}\times$0.19$^{\prime\prime}$, -77.60$^\circ$    & 21.46 & $<$6.43$\times10^{-2}$ & - \\ 
			 & 11.5       &0.23$^{\prime\prime}\times$0.15$^{\prime\prime}$, -77.03$^\circ$    & 24.50 & $<$7.35$\times10^{-2}$ & \\ \hline
B31459+399NED01		 & 8.5        &0.21$^{\prime\prime}\times$0.19$^{\prime\prime}$, -42.04$^\circ$    & 21.97 & 10.66  &-1.04$\pm$0.07  \\
			 & 11.5       &0.16$^{\prime\prime}\times$0.15$^{\prime\prime}$, -45.96$^\circ$    & 24.93 & 7.99   & \\ \hline
2MASX J15042218+4741116  & 8.5        &0.26$^{\prime\prime}\times$0.21$^{\prime\prime}$,  8.94$^\circ$     & 23.76 & 1.81   & -0.11$\pm$0.07 \\
			 & 11.5       &0.19$^{\prime\prime}\times$0.16$^{\prime\prime}$, -5.77$^\circ$     & 27.34 & 1.75   & \\ \hline
SDSS J160027.78+083743.0 & 8.5        &0.23$^{\prime\prime}\times$0.19$^{\prime\prime}$, -53.99$^\circ$    & 20.55 & 0.75   & 0.32$\pm$0.40 \\
                         & 11.5       &0.19$^{\prime\prime}\times$0.16$^{\prime\prime}$, -5.77$^\circ$     & 23.79 & 0.83   & \\ \hline   
2MASX J16441390+2528286  & 6.0        &0.43$^{\prime\prime}\times$0.30$^{\prime\prime}$, -69.54$^\circ$    & 14.40 &11.79    & -0.35$\pm$0.02 \\                                                
                         & 15.0       &0.60$^{\prime\prime}\times$0.38$^{\prime\prime}$, -73.07$^\circ$    & 7.76  & 8.93   &  \\ \hline                
2MASX J23044283-0933454  & 6.0        &0.48$^{\prime\prime}\times$0.28$^{\prime\prime}$, 6.60$^\circ$      & 12.85 & 1.60   &-1.03$\pm$0.11\\ 
                         & 11.5       &0.52$^{\prime\prime}\times$0.38$^{\prime\prime}$, 78.14$^\circ$     & 7.76  & 1.16   &  \\\hline
2MASX J233604.04+000447  & 6.0        &0.41$^{\prime\prime}\times$0.29$^{\prime\prime}$, 0.34$^\circ$      & 19.20 & 3.18   &-1.06$\pm$0.05    \\                                                                  
                         & 15.0       &0.58$^{\prime\prime}\times$0.38$^{\prime\prime}$, -5.72$^\circ$     & 7.39  & 1.20   &  \\ \hline

\end{tabular}
}
\end{table*}

\begin{table*}
\centering
\caption{{\bf Radio properties of dual AGN in our sample:} column 1: Name of the dual AGN galaxies; 
column 2: this shows core no. Primary core is called as core 1 and secondary core as core 2; 
column 3: the observed frequencies in GHz unit; column 4: Beamsize; column 5: Flux density (S$_{int}$) in mJy unit;
column 6: noise in $\mu$Jy/beam; column 7: the spectral index value of individual core; 
column 8:separation (D) of the cores in arcsec; column 9: the separation (D) in kpc. {\bf Note:} $^{\star}$The second core (B) in J1006 shows two lobes. The average $\alpha$ is given. The primary core (A) of J1006 and secondary core (B) of J1355 have an upper limit of $\alpha$ as these are detected with 4-5$\sigma$ confidence.}
\label{dual_core}
\resizebox{\textwidth}{!}{
\begin{tabular}{@{}p{40mm}lclllp{20mm}lll}
\hline\hline name                                      & core & Frequency  &  Beamsize                                                        & S$_{int}$ & Noise &\hspace{0.7cm} $\alpha$                  & D      & D     \\
                                        &                       & in GHz    & ($\theta_{1} \times \theta_{2}$, PA)  & (mJy) & ($\mu$Jy/beam) &  & ($^{\prime\prime}$) & (kpc) \\\hline
                                        
\multirow{4}{40mm}{SDSS J100602.13+071130.9} & 1    & 8.5       & 0.28$^{\prime\prime}$ $\times$0.22$^{\prime\prime}$, 87.50$^\circ$ &  $\le$7.44$\times10^{-2}$      & 15.09 & \multirow{2}{20mm}{$\le$$-$0.93$\pm$1.16}  & \multirow{4}{5mm}{5.0} & \multirow{4}{5mm}{12.0} \\
& 1    & 11.5      & 0.19$^{\prime\prime}\times$0.17$^{\prime\prime}$, 52.98$^\circ$ & $\le$5.77$\times10^{-2}$  & 16.52  &                        &                       &                        \\
                                          & 2    & 8.5       & 0.28$^{\prime\prime}\times$0.22$^{\prime\prime}$, 87.50$^\circ$ & 1.01         & 15.09 & \multirow{2}{20mm}{$-$1.19$\pm$0.50$^{\star}$}  &                       &                        \\
                                          & 2    & 11.5      & 0.19$^{\prime\prime}\times$0.17$^{\prime\prime}$, 52.98$^\circ$ & 0.72         & 16.52  &                        &                       &                        \\ \hline
\multirow{4}{40mm}{SDSS J135558.08+001530.6} & 1    & 8.5       & 0.31$^{\prime\prime}\times$0.24$^{\prime\prime}$, -44.09$^\circ$ &  6.00$\times10^{-1}$      & 17.92 & \multirow{2}{20mm}{$-1.18\pm0.53$}  & \multirow{4}{5mm}{3.1}  & \multirow{4}{5mm}{8.2}  \\
                                          & 1    & 11.5      & 0.23$^{\prime\prime}\times$0.18$^{\prime\prime}$, 45.19$^\circ$ & 4.24$\times10^{-1}$        & 20.80 &                        &                       &                        \\
                                          & 2    & 8.5       & 0.31$^{\prime\prime}\times$0.24$^{\prime\prime}$, -44.09$^\circ$ & $\le$1.30$\times10^{-1}$        & 17.92 & \multirow{2}{20mm}{$\le$ $-0.97\pm1.07$}  &                       &                        \\
                                          & 2    & 11.5      & 0.23$^{\prime\prime}\times$0.18$^{\prime\prime}$, 45.19$^\circ$ &  $\le$1.01$\times10^{-1}$       & 20.80 &                        &                       &                       \\  \hline
\multirow{4}{40mm}{2MASX J16170895+2226279}    & 1    & 6.0         & 0.39$^{\prime\prime}\times$0.30$^{\prime\prime}$, -72.25$^\circ$ &  7.89$\times10^{-1}$        & 12.78 & \multirow{2}{20mm}{$-0.95\pm0.10$}  & \multirow{4}{5mm}{4.3}  & \multirow{4}{5mm}{5.6}  \\
                                          & 1    & 15.0        & 0.59$^{\prime\prime}\times$0.39$^{\prime\prime}$, -71.76$^\circ$& 3.29$\times10^{-1}$        & 7.05  &                        &                       &                        \\
                                          & 2    & 6.0         & 0.39$^{\prime\prime}\times$0.30$^{\prime\prime}$, -72.25$^\circ$ & 3.14$\times10^{-1}$        & 12.78 & \multirow{2}{20mm}{$-0.28\pm0.14$} &                       &                        \\
                                          & 2    & 15.0        & 0.59$^{\prime\prime}\times$0.39$^{\prime\prime}$, -71.76$^\circ$ & 2.37$\times10^{-1}$        & 7.05  &                        &                       &                        \\ \hline
\end{tabular}
}
\end{table*}

\begin{table*}
\centering
\caption{{\bf Radio properties of sample galaxies from FIRST (1.4 GHz) map:} Column 1: SDSS name of the galaxy; Column 2: Integrated flux density (S$_{int}$) of the FIRST image in Jy unit; Column 3: Peak intensity (S$_{peak}$) of the FIRST image in Jy/beam; Column 4: The ratio of the integrated flux density to the peak intensity ($\theta$)=(S$_{int}$/S$_{peak})^{1/2}$.}
\label{gal_first}
\resizebox{\textwidth}{!}{
\begin{tabular}{ccccccc}
\hline Name        &  \begin{tabular}[c]{@{}c@{}}FIRST integrated flux density\\ (Jy)\end{tabular} &  \begin{tabular}[c]{@{}c@{}}FIRST peak intensity\\ (Jy~beam$^{-1}$)\end{tabular}  & $\theta$ =(S$_{int}$/S$_{peak})^{1/2}$\\ \hline

2MASXJ09120164+5320369    &  0.13                  &0.12                   &1.04\\
UGC 05353                 &  3.15$\times 10^{-3}$  &3.14$\times 10^{-3}$   &1.00\\
SDSS J100602.13+071130.9  &  7.80$\times 10^{-3}$  &6.46$\times 10^{-3}$   &1.09\\
SDSS J110215.68+290725.2  &  Core-jet               & $-$                    & $-$\\
SDSS J132318.81+030807.1  &  Core-jet               & $-$                    & $-$\\
2MASX J13245059+1758152   &  1.41$\times 10^{-2}$  &1.38$\times 10^{-2}$   &1.01\\
2MASX J13490964+0404487   &  6.28$\times 10^{-2}$  & 5.96$\times 10^{-2}$  &1.02\\
SDSS J135558.08+001530.6  &  4.64$\times 10^{-3}$  &3.88$\times 10^{-3}$   &1.09\\
2MASX J14131625+2119374   &  2.49$\times 10^{-3}$  &2.19$\times 10^{-3}$   &1.06\\
2MASX J14203147+4008166   &  7.07$\times10^{-3}$   &6.57$\times10^{-3}$    &1.03\\
2MASXJ14454130+3341080    &  4.21$\times10^{-3}$   &3.87$\times10^{-3}$    &1.04\\
2MASX J15001769+1051100   &  8.88$\times10^{-3}$   & 2.56$\times10^{-3}$   &1.86\\
B31459+399NED01           &  4.56$\times10^{-2}$   &4.26$\times10^{-2}$    &1.03\\
2MASX J15042218+4741116   &  9.45$\times10^{-3}$   & 5.88$\times10^{-3}$   &1.26\\
SDSS J160027+083742       &  hotspot                & $-$                    & $-$\\
2MASX J16170895+2226279   &  4.21$\times10^{-3}$   &2.81$\times10^{-3}$    &1.22\\
2MASX J16441390+2528286   &  7.94$\times10^{-3}$   &7.56$\times10^{-3}$    &1.02\\
2MASX J23044283-0933454   &  8.78$\times10^{-3}$   & 8.66$\times10^{-3}$   &1.00\\
SDSS J233604.04+000447.1  &  1.60$\times10^{-2}$   &1.38$\times10^{-2}$    &1.07\\\hline

\end{tabular}
}
\end{table*}

\begin{table*}
\centering
\caption{{\bf Mass of the BH (M$_{BH}$), Eddington ratio ($\lambda$) and Star-formation rate (SFR) calculations:} Column 1: The name of the galaxies; 
Column 2: SMBH mass in M$_\odot$ units using the M-${\sigma}_*$ relation where $\sigma$ is taken from Column 9, Table \ref{sam_gal}; 
Column 3: The [O III] luminosity in erg~s$^{-1}$ units; Column 4: the bolometric luminosity in erg~s$^{-1}$ units; 
Column 5: The Eddington luminosity in erg~s$^{-1}$ units calculated using M$_{BH}$ (Column 2);
Column 6: The Eddington ratio; 
Column 7: H${\alpha}$ luminosity (L$_{H\alpha}$) in erg s$^{-1}$ units; 
Column 8: Radio luminosity in W~Hz$^{-1}$ units;
Column 9: The star formation rate (SFR$_{H\alpha}$) in M$_\odot$ yr$^{-1}$ using H${\alpha}$ flux from SDSS spectra;
Column 10: The star formation rate (SFR$_{Radio}$) in M$_\odot$ yr$^{-1}$ using radio luminosity at 1.4 GHz from NVSS survey;}
\label{gal_prop}
\resizebox{\textwidth}{!}{
\begin{tabular}{lccccccrcc}
\hline \hline 
Name                    & \begin{tabular}[c]{@{}c@{}}M$_{BH}$\\ (M$_\odot$)\end{tabular}    & \begin{tabular}[c]{@{}c@{}}L$_{O[III]}$\\(erg s$^{-1}$)\end{tabular} 
& \begin{tabular}[c]{@{}c@{}}L$_{Bol}$\\ (erg s$^{-1}$)\end{tabular} & \begin{tabular}[c]{@{}c@{}}L$_{Edd}$\\(erg s$^{-1}$)\end{tabular} & $\lambda=$ L$_{Bol}$/L$_{Edd}$ 
& \begin{tabular}[c]{@{}c@{}}L$_{H\alpha}$\\ (erg s$^{-1}$)\end{tabular} & \begin{tabular}[c]{@{}c@{}}L$_{Radio}$\\ (W~Hz$^{-1}$)\end{tabular}
& \begin{tabular}[c]{@{}c@{}} SFR$_{H\alpha}$\\(M$_\odot$ yr$^{-1}$)\end{tabular}  & \begin{tabular}[c]{@{}c@{}}SFR$_{Radio}$\\(M$_\odot$ yr$^{-1}$)\end{tabular}  \\\hline\hline

2MASX J09120164+5320369  &$(2.65 \pm 0.68)\times$10$^8$   &$(6.30 \pm 0.63)\times$10$^{40}$    &$(2.20 \pm 0.22)\times$10$^{44}$     &$(3.18 \pm 0.82)\times$10$^{46}$    &$(0.06 \pm 0.01)\times$10$^{-1}$   &$(5.65 \pm 0.30)\times$10$^{40}$    &$(4.19 \pm 0.01)\times$10$^{24}$           & 0.44 $\pm$ 0.02    &  310.67 $\pm$ 3.16\\
UGC 05353                &$(2.33 \pm 0.26)\times$10$^9$   &$(1.02 \pm 0.25)\times$10$^{39}$    &$(3.60 \pm 0.88)\times$10$^{42}$     &$(2.80 \pm 0.31)\times$10$^{47}$    &$(1.28 \pm 0.34)\times$10$^{-5}$   &$(1.63 \pm 0.31)\times$10$^{40}$    &$(1.82 \pm 0.47)\times$10$^{21}$           & 0.12 $\pm$ 0.02    &  0.13 $\pm$ 0.11\\
SDSS J100602.13+071130.9 &$(7.40 \pm 3.73)\times$10$^7$   &$(2.69 \pm 0.03)\times$10$^{41}$    &$(9.41 \pm 0.13)\times$10$^{44}$     &$(8.88 \pm 4.48)\times$10$^{45}$    &(0.10 $\pm$ 0.05)                  &$(4.83 \pm 0.03)\times$10$^{41}$    & ---           			       & 3.81 $\pm$ 0.03    &  ---            \\ 
SDSS J110215.68+290725.2 &$(3.15 \pm 0.66)\times$10$^8$   &$(1.29 \pm 0.05)\times$10$^{41}$    &$(4.51 \pm 0.18)\times$10$^{44}$     &$(3.78 \pm 0.79)\times$10$^{46}$    &$(0.11 \pm 0.02)\times$10$^{-1}$   &$(1.01 \pm 0.04)\times$10$^{41}$    & ---         			       & 0.79 $\pm$ 0.03    &  ---        \\
SDSS J132318.81+030807.1 &$(3.03 \pm 1.69)\times$10$^8$   &$(1.56 \pm 0.02)\times$10$^{42}$    &$(5.48 \pm 0.09)\times$10$^{45}$     &$(3.64 \pm 2.02)\times$10$^{46}$    &(0.15 $\pm$ 0.08)                  &$(6.31 \pm 0.18)\times$10$^{41}$    & ---          			       & 4.98 $\pm$ 0.14    &   ----       \\
2MASX J13245059+1758152  &$(5.03 \pm 1.97)\times$10$^6$   &$(1.43 \pm 0.10)\times$10$^{40}$    &$(5.03 \pm 0.37)\times$10$^{43}$     &$(6.03 \pm 2.37)\times$10$^{44}$    &(0.08 $\pm$ 0.03)                  &$(9.06 \pm 0.54)\times$10$^{40}$    &$(2.33 \pm 0.07)\times$10$^{23}$           & 0.71 $\pm$ 0.04    &   17.24 $\pm$ 1.81\\
2MASX J13490964+0404487  &$(2.48 \pm 0.42)\times$10$^8$   &$(1.50 \pm 0.32)\times$10$^{40}$    &$(5.25 \pm 1.12)\times$10$^{43}$     &$(2.98 \pm 0.51)\times$10$^{46}$    &$(0.17 \pm 0.04)\times$10$^{-2}$   &$(2.19 \pm 0.27)\times$10$^{40}$    &$(9.32 \pm 0.07)\times$10$^{23}$           & 0.17 $\pm$ 0.02    &   69.03 $\pm$ 1.79\\
SDSS J135558.08+001530.6 &$(1.53 \pm 1.05)\times$10$^9$   &$(2.98 \pm 0.23)\times$10$^{40}$    &$(1.04 \pm 0.08)\times$10$^{44}$     &$(1.84 \pm 1.27)\times$10$^{47}$    &$(0.05 \pm 0.03)\times$10$^{-2}$   &$(1.02 \pm 0.08)\times$10$^{41}$    & ---          			       & 0.80 $\pm$ 0.06    &    ---      \\
2MASX J14131625+2119374  &$(5.45 \pm 1.66)\times$10$^8$   &$(9.89 \pm 0.94)\times$10$^{40}$    &$(3.46 \pm 0.32)\times$10$^{44}$     &$(6.54 \pm 1.99)\times$10$^{46}$    &$(0.05 \pm 0.01)\times$10$^{-1}$   &$(6.49 \pm 0.66)\times$10$^{40}$    & ---          			       & 0.51 $\pm$ 0.05    &      ---    \\
2MASX J14203147+4008166  &$(4.94 \pm 1.87)\times$10$^8$   &$(6.26 \pm 1.12)\times$10$^{40}$    &$(2.19 \pm 0.39)\times$10$^{44}$     &$(5.93 \pm 2.25)\times$10$^{46}$    &$(0.03 \pm 0.01)\times$10$^{-1}$   &$(6.89 \pm 1.34)\times$10$^{40}$    &$(1.66 \pm 0.05)\times$10$^{23}$           & 0.54 $\pm$ 0.10    &   122.68 $\pm$ 13.85   \\
2MASX J14454130+3341080  &$(8.92 \pm 3.48)\times$10$^7$   &$(7.64 \pm 0.39)\times$10$^{41}$    &$(2.67 \pm 0.13)\times$10$^{45}$     &$(1.07 \pm 0.41)\times$10$^{46}$    &(0.24$\pm$ 0.09)                   &$(4.24 \pm 0.12)\times$10$^{41}$    &$(1.85 \pm 0.17)\times$10$^{23}$           & 3.35 $\pm$ 0.09    &   13.71 $\pm$  4.34    \\
2MASX J15001769+1051100  &$(2.08 \pm 0.58)\times$10$^8$   &$(3.67 \pm 0.76)\times$10$^{40}$    &$(1.28 \pm 0.26)\times$10$^{44}$     &$(2.50 \pm 0.69)\times$10$^{46}$    &$(0.05 \pm 0.01)\times$10$^{-1}$   &$(1.19 \pm 0.16)\times$10$^{41}$    &$(7.87 \pm 0.35)\times$10$^{23}$           & 0.93 $\pm$ 0.12    &   58.27 $\pm$  8.79    \\
B31459+399NED01          &$(6.75 \pm 5.35)\times$10$^7$   &$(1.93 \pm 0.19)\times$10$^{41}$    &$(6.76 \pm 0.67)\times$10$^{44}$     &$(8.10 \pm 6.42)\times$10$^{45}$    &$(0.83 \pm 0.66)\times$10$^{-1}$   &$(9.76 \pm 6.63)\times$10$^{40}$    &$(1.81 \pm 0.02)\times$10$^{25}$           & 0.77 $\pm$ 0.52    &   1339.18 $\pm$ 70.21   \\
2MASX J15042218+4741116  &$(6.52 \pm 2.89)\times$10$^7$   &$(5.10 \pm 0.73)\times$10$^{39}$    &$(1.78 \pm 0.27)\times$10$^{43}$     &$(7.82 \pm 3.47)\times$10$^{45}$    &$(0.02 \pm 0.01)\times$10$^{-1}$   &$(3.25 \pm 0.50)\times$10$^{40}$    &$(2.18 \pm 0.07)\times$10$^{23}$           & 0.25 $\pm$ 0.04    &   16.18 $\pm$ 1.75\\  
SDSS J160027.78+083743.0 &$(2.87 \pm 1.01)\times$10$^8$   &$(3.74 \pm 0.09)\times$10$^{41}$    &$(1.30 \pm 0.03)\times$10$^{45}$     &$(3.44 \pm 1.22)\times$10$^{46}$    &(0.03 $\pm$ 0.01)                  &$(2.11 \pm 0.08)\times$10$^{41}$    & ---       				       & 1.67 $\pm$ 0.06    &    ---\\
2MASX J16170895+2226279  &$(2.24 \pm 1.03)\times$10$^7$   &$(7.54 \pm 0.24)\times$10$^{40}$    &$(2.64 \pm 0.08)\times$10$^{44}$     &$(2.69 \pm 1.24)\times$10$^{45}$    &(0.09 $\pm$ 0.04)                  &$(5.38 \pm 0.17)\times$10$^{40}$    &$(2.40 \pm 0.31)\times$10$^{22}$           & 0.42 $\pm$ 0.01    &   1.78 $\pm$  0.77  \\
2MASX J16441390+2528286  &$(6.28 \pm 1.00)\times$10$^8$   &$(1.13 \pm 0.15)\times$10$^{40}$    &$(3.96 \pm 0.52)\times$10$^{43}$     &$(7.53 \pm 1.20)\times$10$^{46}$    &$(0.05 \pm 0.01)\times$10$^{-2}$   &$(5.10 \pm 3.34)\times$10$^{39}$    &$(5.92 \pm 0.31)\times$10$^{22}$           & 0.04 $\pm$ 0.02    &   4.38 $\pm$ 0.76  \\
2MASX J23044283-0933454  &$(2.68 \pm 0.42)\times$10$^7$   &$(1.65 \pm 0.07)\times$10$^{40}$    &$(5.78 \pm 0.27)\times$10$^{43}$     &$(3.22 \pm 0.50)\times$10$^{45}$    &$(0.17 \pm 0.02)\times$10$^{-1}$   &$(1.75 \pm 0.05)\times$10$^{40}$    &$(2.24 \pm 0.11)\times$10$^{22}$           & 0.13 $\pm$ 0.01    &   1.66  $\pm$ 0.27  \\
2MASX J233604.04+000447  &$(7.35 \pm 3.06)\times$10$^7$   &$(3.81 \pm 0.10)\times$10$^{40}$    &$(1.33 \pm 0.36)\times$10$^{44}$     &$(8.82 \pm 3.63)\times$10$^{45}$    &$(0.15 \pm 0.06)\times$10$^{-1}$   &$(3.69 \pm 0.14)\times$10$^{40}$    &$(2.46 \pm 0.06)\times$10$^{23}$           & 0.29 $\pm$ 0.01    &   18.20 $\pm$ 1.53   \\ \hline 

\end{tabular}
}
\end{table*}

\end{document}